\documentclass[fleqn,usenatbib]{mnras}
\usepackage{newtxtext,newtxmath}
\usepackage[T1]{fontenc}
\usepackage{units}
\usepackage{lineno}
\usepackage{color}
\usepackage{epsfig}
\usepackage{float}
\usepackage{bm}
\DeclareRobustCommand{\VAN}[3]{#2}
\let\VANthebibliography\thebibliography
\def\thebibliography{\DeclareRobustCommand{\VAN}[3]{##3}\VANthebibliography}
\usepackage{graphicx}
\usepackage{amsmath} 	 

\usepackage{amssymb}	 
\usepackage{amsfonts}
\usepackage{caption}
\usepackage{subfigure}
\usepackage{multirow}
\usepackage[figuresright]{rotating}

\def\be{\begin{equation}}
\def\ee{\end{equation}}

\def\bes{\begin{align}}
\def\ees{\end{align}}

\newcommand{\fraction}[3]{\left(\frac{#1}{#2}\right)^{#3}}
\newcommand{\fractionz}[2]{\left(\frac{#1}{#2}\right)} 

\title[RM Variations of Repeating FRB Sources]{Faraday Rotation Measure Variations of Repeating Fast Radio Burst Sources} 

\author[Yang et al.]{
Yuan-Pei Yang$^{1,2}$\thanks{E-mail: ypyang@ynu.edu.cn (YPY)},
Siyao Xu$^{3}$\thanks{E-mail: sxu@ias.edu (SX)}
and Bing Zhang $^{4,5}$ \thanks{E-mail: bing.zhang@unlv.edu (BZ)
}
\\
$^{1}$South-Western Institute for Astronomy Research, Yunnan University, Kunming, Yunnan 650504, China\\
$^{2}$Purple Mountain Observatory, Chinese Academy of Sciences, Nanjing 210023, China\\
$^{3}$Institute for Advanced Study, 1 Einstein Drive, Princeton, NJ 08540, USA\\
$^{4}$Nevada Center for Astrophysics, University of Nevada, Las Vegas, NV 89154, USA\\
$^{5}$Department of Physics and Astronomy, University of Nevada, Las Vegas, NV 89154, USA\\
}

\date{Accepted XXX. Received YYY; in original form ZZZ}

\pubyear{2022}

\begin{document}
\label{firstpage}
\pagerange{\pageref{firstpage}--\pageref{lastpage}}
\maketitle

\begin{abstract}
Recently, some fast radio burst (FRB) repeaters were reported to exhibit complex, diverse variations of Faraday rotation measures (RMs), which implies that they are surrounded by an inhomogeneous, dynamically evolving, magnetized environment. We systematically investigate some possible astrophysical processes that may cause RM variations of an FRB repeater. The processes include (1) a supernova remnant (SNR) with a fluctuating medium; (2) a binary system with stellar winds from a massive/giant star companion or stellar flares from a low-mass star companion; (3) a pair plasma medium from a neutron star (including pulsar winds, pulsar wind nebulae, and magnetar flares); (4) outflows from a massive black hole. For the SNR scenario, a large relative RM variation within a few years requires that the SNR is young with a thin and local anisotropic shell, or the size of dense gas clouds in interstellar/circumstellar medium around the SNR is extremely small. If the RM variation is caused by the companion medium in a binary system, it is more likely from the stellar winds of a massive/giant star companion. The RM variation contributed by stellar flares from a low-mass star is disfavored, because this scenario predicts an extremely large relative RM variation during a short period of time. The scenarios invoking a pair plasma from a neutron star can be ruled out due to their extremely low RM contributions. Outflows from a massive black hole could provide a large RM variation if the FRB source is in the vicinity of the black hole.

\end{abstract} 

\begin{keywords}

(transients:) fast radio bursts -- (stars:) pulsars:general -- radio continuum: transients -- ISM: structure
\end{keywords}

\section{Introduction} 

Fast radio bursts (FRBs) are mysterious radio transients with millisecond durations and extremely high brightness temperatures at cosmological distances. So far, over 600 FRB sources have been detected, dozens of which exhibited a repeating behavior \citep[e.g.,][]{CHIME21}. 
However, their physical origin is still not well understood due to the complexity and diversity of the observations \citep[e.g.,][]{Cordes19,Zhang20,Xiao21}. For example, a Galactic FRB, FRB 200428, was detected to be associated with the magnetar SGR J1935+2154 \citep{Bochenek20,CHIME20,Mereghetti20,Li20,Ridnaia20,Tavani20}, implying that at least some FRBs originate from magnetars born from the core collapse of massive stars \citep{Katz16,Murase16,Beloborodov17,Kumar17,Yang18,Yang21,Metzger19,Wadiasingh19,Lu20,Margalit20,Zhang22,Wang22b,Qu22}. 
However, such a magnetar formation is challenged by the observation of another nearby FRB, FRB 20200120E, which was localized to be in a globular cluster of the nearby galaxy M81 \citep{Bhardwaj21,Kirsten22}. The extremely old age of the globular cluster implies that it is more likely produced by an old object or a system associated with a compact binary merger \citep{zhang20c,Kremer21,Lu22}.
Therefore, multiple physical origins for the FRB population seem increasingly likely.

In addition to the FRB sources themselves, propagating effects, e.g., dispersion, Faraday rotation, temporal scattering, scintillation, depolarization, and gravitational/plasma lensing, also play important roles to interpret FRB observations and constrain the properties of the FRB environment \citep[e.g.,][]{Xu16,Cordes17,Yang17,Li18b,Yang20b,Er20,Beniamini22,Yang22,Kumar22}. 
Dispersion measure (DM) and Faraday rotation measure (RM) are two of the most important measurable quantities for FRBs. For a repeating FRB source, the variations of its DM and RM would provide clues to study the properties of its near-source plasma. \cite{Yang17} studied various possible origins to cause DM variations of an FRB repeater and concluded that the plasma local to the FRB source is the most likely cause.
Different from DM which usually has a significant contribution from the intergalactic medium, the observed RM of an FRB, especially when it is large, is likely contributed by 
a highly magnetized environment near the FRB source, because the RM contribution from the intergalactic medium is very small with a typical value of $|{\rm RM_{IGM}}|\ll 10~{\rm rad~m^{-2}}$ \citep{Akahori16} and because the contribution from the interstellar medium in the Milky Way is usually $|{\rm RM_{MW}}|\lesssim100~{\rm rad~m^{-2}}$ at high latitudes \citep{Hutschenreuter22}.
The first known repeater, FRB 121102, shows the largest RM of $|{\rm RM}|\sim10^5~{\rm rad~m^{-2}}$ among all observed FRB sources \citep{Michilli18}, which it decreased by $\sim30\%$ during one year \citep{Hilmarsson21}. This may be caused by
the expansion of a young supernova remnant (SNR) \citep{Piro18}, a magnetar nebula \citep{Margalit18}, or an ejecta from a compact binary merger \citep{Zhao21} around the FRB source, although other scenarios  (see discussion below) may also be possible. Another active repeater, FRB 190520B, has an extremely large host DM with ${\rm DM_{host}}\sim900~{\rm pc~cm^{-3}}$ \citep{Niu22}, which is nearly an order of magnitude higher than those of other FRBs. Meanwhile, its RM value reaches $\sim10^4~{\rm rad~m^{-2}}$ that is second in line next to FRB 121102, and there appears an RM sign reversal within a few months \citep{Anna-Thomas22,Dai22}. Such a large RM and significant RM reversal directly suggest that the FRB environment is magnetized and dynamically evolving. 
It is worth noting that both FRB 121102 and FRB 190520B 
are associated with a compact persistent radio source with a wide emission spectrum \citep{Chatterjee17,Niu22}, which implies that a persistent radio source and a large RM are likely physically related to each other \citep{Yang20a,Yang22}.

Other FRB repeaters also show complex and diverse RM variations. FRB 20201124A showed an irregular RM variation over one month. Some bursts appeared to have circular polarization and frequency-dependent oscillating polarization properties \citep{Xu21}. FRB 180916B with a 16.33-day periodic activity \citep{CHIME20b} exhibited RM variation with a stochastic component and a secular component \citep{Mckinven22}. 
In summary, RM variations seem to be a common feature for all FRB repeaters.

Such significant RM variations suggest that repeating FRBs are likely surrounded by an inhomogeneous and dynamically evolving Faraday screen. When a radio burst propagates in the screen, 
two important effects are involved: 1) the radio bursts would be depolarized due to different RMs at different paths. Very recently, \citet{Feng22} reported that active FRB repeaters exhibit conspicuous frequency-dependent depolarization that can be well described by the multi-path propagation effect in an inhomogeneous magnetized plasma \citep{Yang22}. 2) a significant circular polarization would be generated due to the superposition of electromagnetic waves with different phases and polarization angles from different paths \citep{Beniamini22}.

In this paper, we investigate the possible physical mechanisms that may cause RM variations from a repeating FRB source, and discuss the physical implications of the observed RM variations. The paper is organized as follows. We discuss the necessary conditions and theoretical implications of Faraday rotation in Section \ref{general}. The physical origins of random and secular RM evolution are generally analyzed in Section \ref{FaradayScreen}, where 
the random RM variation is caused by the relative motion between the FRB source and a Faraday screen with an inhomogeneous magnetized medium, and the secular RM evolution may be caused by an expanding shell or the orbital motion in a binary system. 
In Section \ref{scenarios}, we discuss different astrophysical scenarios, including SNRs with an inhomogeneous medium in Section \ref{SNR}, stellar winds from a massive/giant star companion in Section \ref{stellarwind}, stellar flares from a low-mass star companion in Section \ref{stellarflare}, pulsar winds, pulsar wind nebulae and magnetar flares in Section \ref{pair}, and magnetized outflows from a massive black hole in Section \ref{BH}). We discuss the observed properties and implications of some specific FRB repeaters in Section \ref{discussion}. The results are summarized in Section \ref{conclusion}. Some detailed calculations are presented in the Appendices.

\section{Rotation measure: a general discussion}\label{general}

In general, the observed Faraday rotation measure, $\rm RM$, is measured by the frequency(wavelength)-dependent polarization angle of linearly polarized waves\footnote{This applies to the one single RM component scenario, in which pure Faraday rotation occurs only for one foreground magneto-ionic plasma, Multiple RM components can cause the observed polarization angle $\psi$ departs from a linear relation with $\lambda^2$, but the RM components could be recovered involving the observed frequency-dependent linear polarization degree if $\psi\propto\lambda^2$ is satisfied in each component \citep{OSullivan12}. }
\be
\psi={\rm RM}\lambda^2,\label{psiwavelength}
\ee
where $\psi$ is the polarization angle of the electromagnetic wave of wavelength $\lambda$ with respect to that of infinite frequency. The necessary conditions to measure ${\rm RM}$ of a source include: 1) the electromagnetic waves must contain a significant linear polarization component; 2) the polarization angle must satisfy $\psi\propto\lambda^2$.

We should note that the condition for $\psi\propto\lambda^2$ to be relevant requires that $\omega\gg\max(\omega_B,\omega_p)$ is satisfied according to the dispersion relation of circularly polarized waves, where $\omega_B=eB/m_ec$ is the cyclotron frequency, and $\omega_p=(4\pi e^2n_e/m_e)^{1/2}$ is the plasma frequency. Because $\omega>\omega_p$ is usually satisfied in most meaningful scenarios, the necessary condition for $\psi\propto\lambda^2$ can be translated to a constraint on the magnetic field strength, i.e.
\be
B\ll B_c=\frac{2\pi m_ec\nu}{e}\simeq360~{\rm G}\fractionz{\nu}{1~{\rm GHz}},\label{Bc}
\ee
according to $\omega\gg\omega_B$.
Therefore, although the FRB engine (e.g. a neutron star or a black hole) has a strong magnetic field near the engine, the region with $B>B_c$ cannot contribute to the observed RM\footnote{The medium near the engine (e.g. a neutron star) is likely a relativistic pair plasma. The RM contribution by such a pair plasma is not important anyway, see the discussion in Section \ref{pair} and Appendix \ref{RMpair}.}. 

The observed RM of an extragalactic FRB can be decomposed in terms of the contributions of various plasma components along the line of sight, i.e.
\be
{\rm RM_{obs}}={\rm RM_{\rm ion}}+{\rm RM_{MW}}+{\rm RM_{IGM}}+\frac{{\rm RM_{host}}}{(1+z)^2}+\frac{{\rm RM_{loc}}}{(1+z)^2},
\ee
where $z$ is the redshift of the host galaxy, ${\rm RM_{\rm ion}}$ is the contribution from the Earth ionosphere, which is of the order of $|{\rm RM_{\rm ion}}|\sim(0.1-1)~{\rm rad~m^{-2}}$  \citep{Mevius18,Mckinven22},
${\rm RM_{MW}}$ is the contribution from the interstellar medium in the Milky Way, which has a typical absolute value $|{\rm RM_{MW}}|\lesssim100~{\rm rad~m^{-2}}$ at high latitudes \citep{Hutschenreuter22}, 
${\rm RM_{IGM}}$ is the contribution from  the intergalactic medium which has a very small value of $|{\rm RM_{IGM}}|< 10~{\rm rad~m^{-2}}$ \citep{Akahori16}, ${\rm RM_{host}}$ is the contribution from the interstellar medium in the FRB host galaxy, which might be of the same order of magnitude as the Milky Way, and ${\rm RM_{loc}}$ is the contribution from the local plasma near the FRB source. Therefore, an observed RM with a large absolute value of $|{\rm RM_{obs}}|\gtrsim10^3~{\rm rad~m^{-2}}$, e.g., the RMs of FRB 121102 and FRB 190520B \citep{Michilli18,Anna-Thomas22,Dai22}, is expected to be mainly contributed by the local plasma ${\rm RM_{loc}}$. 
Since both the intergalactic medium and interstellar medium cannot have short-term evolution \citep{Yang17}, the observed significant RM variations can only be attributed to the local plasma. In the following discussion, we are only interested in ${\rm RM_{loc}}$, and hereafter directly use the symbol ${\rm RM}$ to denote ${\rm RM_{loc}}$.

\begin{figure}
    \centering
	\includegraphics[width = 1.0\linewidth, trim = 0 50 0 50, clip]{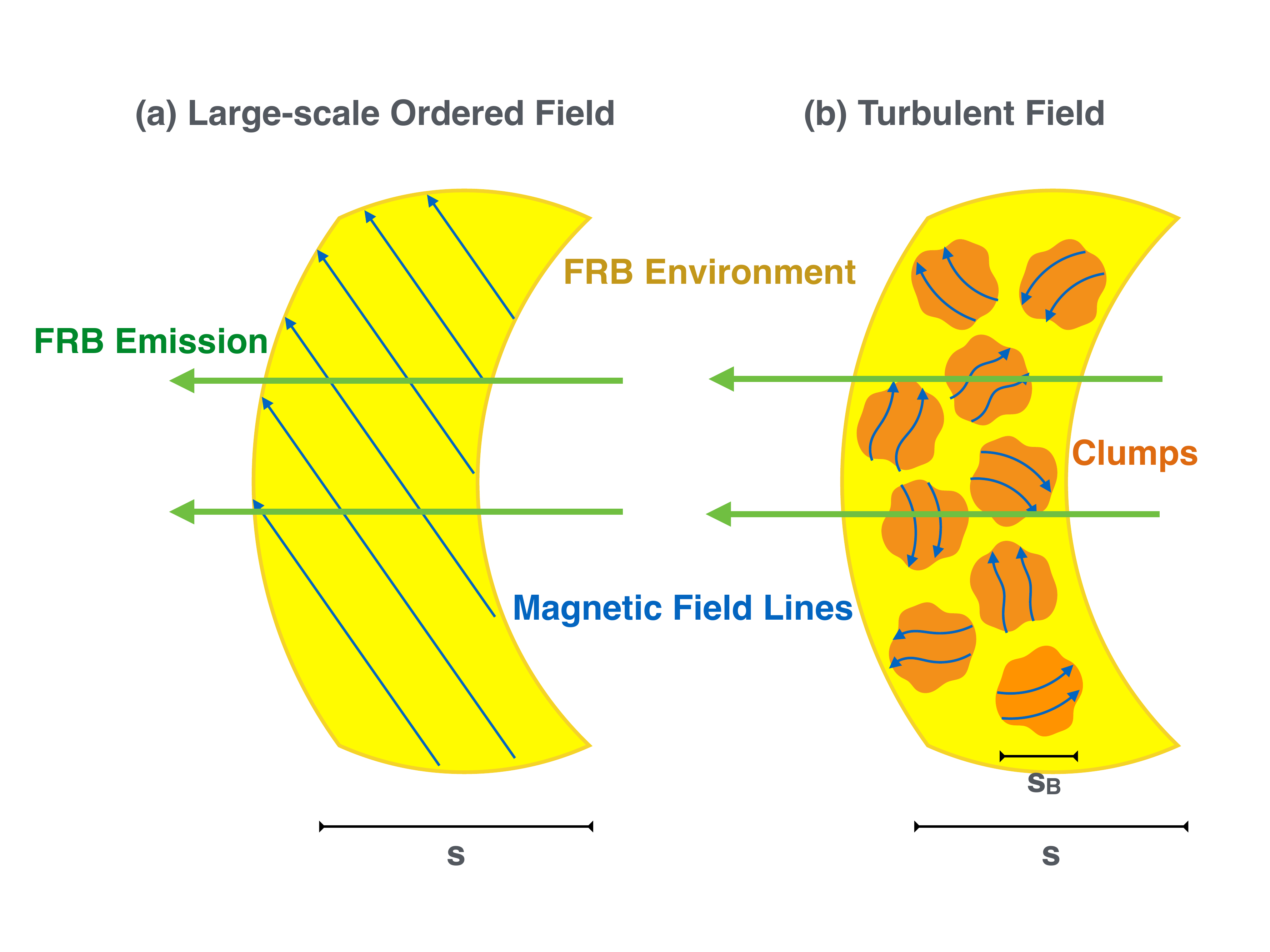}
    \caption{Schematic configurations of a magnetized FRB environment. There are two extreme scenarios: (a) the magnetic fields in the environment are large-scale ordered. (2) the magnetic fields in the environment are turbulent. The yellow regions correspond to the FRB environment. The orange regions correspond to the clumps in a turbulent medium. The blue lines correspond to magnetic field lines.}\label{figfield} 
\end{figure}

For a non-relativistic magneto-ionic (ions+electrons) cold plasma with magnetic field $B$ and electron density $n_e$, the RM could be calculated by the dispersion relation of the electromagnetic wave, and the classical result is
\begin{align}
{\rm RM}&=\frac{e^3}{2\pi m_e^2c^4}\int n_eB_\parallel ds \nonumber\\
&\sim0.81~{\rm rad~m^{-2}}\fractionz{\left<n_e\right>_L}{1~{\rm cm^{-3}}}\fractionz{\left<B_\parallel\right>_L}{1~{\rm \mu G}}\fractionz{s}{1~{\rm pc}} \nonumber\\
&\sim0.81~{\rm rad~m^{-2}}\fractionz{\left<B_\parallel\right>_L}{1~{\rm \mu G}}\fractionz{{\rm DM}}{1~{\rm pc~cm^{-3}}},\label{rm}
\end{align}
where $\left<...\right>_L$ denotes to the average along the line of sight in the local plasma near an FRB source. 
Although there is usually a component of the field that is spatially coherent at the scale of a certain object, the field lines are disordered and there are magnetic fluctuations over scales spanning orders of magnitude due to the presence of turbulence. Thus, both a large-scale ordered magnetic field and a turbulent magnetic field are involved in various astrophysical scenarios, as shown in Figure \ref{figfield}.
The average of $B_\parallel$ depends on the geometric configuration of the magnetic fields along the line of sight.
For example, we consider that the electron density and magnetic field strength are approximately uniform but the magnetic geometry might change along the line of sight.
If the magnetic field is ordered in a large scale (see panel (a) of Figure \ref{figfield}), the average parallel magnetic field $\left<B_\parallel\right>$ would be of the order of the local parallel magnetic field $B_\parallel$, i.e., $\left<B_\parallel\right>_L\sim B_\parallel$. However, if the magnetic field is totally turbulent with a coherent lengthscale of $s_B$ (in a region with lengthscale $s_B$, the field could be treated as approximately uniform with a certain direction, see panel (b) of Figure \ref{figfield}), the average parallel magnetic field would be $\left<B_\parallel\right>_L\sim (s_B/s)^{1/2} B_\parallel$ due to the Poisson r.m.s. fluctuations of the polarization angle \citep[e.g.,][]{Beniamini22,Yang22}. In general, for a local plasma one may write
\begin{align}
\left<B_\parallel\right>_L&\simeq1.23~{\rm \mu G}\fractionz{{\rm RM}}{1~{\rm rad~m^{-2}}}\fraction{{\rm DM}}{1~{\rm pc~cm^{-3}}}{-1}\nonumber\\
&\sim
\left\{
\begin{aligned}
&B_\parallel,&&\text{for ordered field},\\
&(s_B/s)^{1/2}B_\parallel,&&\text{for turbulent fields}. 
\end{aligned}
\right.\label{Bfield}
\end{align}
The above equation has widely been used to estimate the strength of magnetic fields in a region using RM and DM, provided that the RM and DM originate from the same region. 

Some authors \citep[e.g.,][]{Katz21} estimated the magnetic field strength of the local plasma by assuming that the variations of RMs and DMs originate from the same region, so that
$\left<B_\parallel\right>_L\simeq1.23~{\rm \mu G}(\delta{\rm RM}/1~{\rm rad~m^{-2}})(\delta{\rm DM}/1~{\rm pc~cm^{-3}})^{-1}$.
This formula can be easily applied to the observational data, because both $\delta {\rm RM}$ and $\delta {\rm DM}$ are measurable quantities for a repeating source and the variation from the ISM and IGM could be negligible for both DM and RM during the observing time. 
However, such a formula is inapplicable for the scenario that the RM variation is caused by magnetic configuration variation, which is likely a cause of RM variations and has been confirmed by the RM reversal observed in FRB 190520B \citep{Anna-Thomas22,Dai22}. 
For example, let us consider a case that the electron density and magnetic field strength is unchanged but the magnetic field configuration is changing. This gives $\delta {\rm DM}\sim 0$ and $|\delta {\rm RM}|>0$, leading to the unphysical conclusion $\left<B_\parallel\right>_L\rightarrow\infty$ with the formula.
The correct treatment should be to take
differentiation of Eq.(\ref{rm}) and obtain 
\be
\frac{\delta {\rm RM}}{{\rm RM}}\simeq\frac{\delta {\rm DM}}{{\rm DM}}+\frac{\delta \left<B_\parallel\right>_L}{\left<B_\parallel\right>_L}, 
\label{eq:deltaRM}
\ee
which suggests that the relative RM variation is the sum of the relative variations of DM and the average parallel magnetic field. If the relative DM variation is very small, $\delta {\rm DM}/{\rm DM}\ll1$, the observed RM variation would be dominated by the change of field configuration.

\section{Physical origins of random and secular RM variations}\label{FaradayScreen}

Recent observations show that many FRB repeaters appear to have complex, diverse RM variation patterns. 
For example, FRB 121102 exhibited a non-linear decrease of RM absolute value within a few years with several weeks of fluctuations in a short term
\citep{Chatterjee17,Hilmarsson21}.
In an active cycle, FRB 20201124A showed an irregular RM variation during the first 36 days, which is followed by an almost constant RM during a later 18 days \citep{Xu21}. FRB 180916B showed stochastic, small RM variations
followed by a significant secular increasing component over the nine-month period \citep{Mckinven22}. It seems that observations show both random fluctuations and systematic secular evolution. In the following, we present some general discussions on these two scenarios. 

\subsection{Random RM variations}
For random RM variations, the most likely scenario is that there is a Faraday screen with an inhomogeneous medium near the FRB source, and the relative motion between the FRB source and the screen causes irregular RM variations. 
We assume that the timescale of relative motion is shorter than the dynamic timescale of turbulence, i.e. largest eddy turnover time.
The ubiquitous turbulence in astrophysical plasmas naturally induces fluctuations in density and magnetic fields, and hence, RM fluctuations \citep{Minter96,Vogt2005,Lazarian16,Xu16}.
We consider a Faraday screen with thickness $\Delta R$ and assume a statistical homogeneity of the medium. The ``RM structure function'' is used to represent the mean-squared RM difference between two paths separated by a transverse distance $l$, i.e.,
\be
D_{\rm RM}(\overrightarrow{l})\equiv\left<[{\rm RM}(\overrightarrow{x}+\overrightarrow{l})-{\rm RM}(\overrightarrow{x})]^2\right>,\label{SF}
\ee
where $\left<...\right>$ represents an ensemble average. Physically, the RM structure function depends on the power spectra $P(k)$ of the fluctuations of RM density $(n_e B_\parallel)(\overrightarrow{x})$, where $k=2\pi/l$ is the spatial wavenumber. For simplicity, we assume a power-law distribution of\footnote{In the following discussion, we do not separate the RM density fluctuations into fluctuations arising from electron density and magnetic field. If the fluctuations of electron density and magnetic field have different spectral indexes, the spectral index of RM density $(n_e B_\parallel)(\overrightarrow{x})$ would be dominated by the one with larger relative variation \citep{Xu16}, $\delta X/X$, where $X$ denotes $n_e$ or $B_\parallel$.} $P(k)\propto k^\alpha$ for $2\pi L^{-1}<k<2\pi l_0^{-1}$, where $L$ and $l_0$ correspond to the outer scale and inner scale, respectively. 
The power spectrum with the 3D spectral index $\alpha<-3$ is called a ``steep spectrum'' (e.g., $\alpha=-11/3$ for the Kolmogorov scaling), and the fluctuations are dominated by the large scale at $\sim L$, which corresponds to the energy injection scale of turbulence; 
The power spectrum with $\alpha>-3$ is called a ``shallow spectrum'', and inhomogeneity structures are dominated by small-scale fluctuations near $\sim l_0$, which is the energy dissipation scale of turbulence \citep{Lazarian04,Lazarian06,Lazarian16}. 
Shallow density spectra are commonly seen in cold interstellar phases with supersonic turbulence, where the small-scale density enhancement is caused by turbulence compression \citep{Xu17,Xu20}.
For a shallow spectrum, 
the fluctuations on scales larger than $L$ are
model-dependent and $L$ is most likely the largest scale of the fluid system, $L\sim\Delta R$, 
for turbulence driven within the system.

We define the correlation length scale $l_{\rm RM}$ with a correlation $\kappa^2\left<\delta (n_eB_\parallel)(\overrightarrow{x}+\overrightarrow{l_{\rm RM}})\delta (n_eB_\parallel)(\overrightarrow{x})\right>=\sigma_{\rm RM}^2/2$, where $\sigma_{\rm RM}^2=\kappa^2\left<\delta(n_eB_\parallel)^2\right>$ corresponds to the variance of RM density fluctuations multiplying by $\kappa=e^3/(2\pi m_e^2c^4)$. 
For the steep spectrum ($\alpha<-3$), the correlation scale is $l_{\rm RM}\sim L$;
while, for the shallow spectrum ($\alpha>-3$), the correlation scale is $l_{\rm RM}\sim l_0$ \citep{Lazarian16,Xu16}. 
For most astrophysical scenarios, the Faraday screen is considered to be thick\footnote{Here, the definitions of ``thick'' and ``thin'' of a Faraday screen is based on the relation between the correlation length $l_{\rm RM}$ and the screen thickness $\Delta R$ \citep{Lazarian16}. Since the correlation length $l_{\rm RM}$ cannot exceed the largest scale of a system for turbulence driven within it, in most astrophysical scenarios involving a shell with the fluctuations of magnetic field and density as the Faraday screen, the thick screen condition ($l_{\rm RM}<\Delta R$) is usually satisfied.}, i.e., $\Delta R>l_{\rm RM}$. 
Therefore, the RM structure function of a thick screen could be written as (\cite{Lazarian16,Xu16}; see Appendix \ref{RMSF} for a detailed derivation)
\begin{align}
D_{\rm RM}(l)\sim
\left\{
\begin{aligned}
&\sigma_{\rm RM}^2\Delta R l\fraction{l}{l_{\rm RM}}{-(\alpha+3)},&&l_0<l<l_{\rm RM}\sim L,\\
&\sigma_{\rm RM}^2\Delta R l_{\rm RM},&&l>l_{\rm RM}\sim L. 
\end{aligned}
\right.\label{sf1}
\end{align}
for a steep spectrum ($-4<\alpha<-3$) and $L\sim l_{\rm RM}$, and 
\begin{align}
D_{\rm RM}(l)\sim
\left\{
\begin{aligned}
&\sigma_{\rm RM}^2\Delta R l\fraction{l}{l_{\rm RM}}{-(\alpha+3)},&&l_0\sim l_{\rm RM}<l<L,\\
&\sigma_{\rm RM}^2\Delta R^2\fraction{\Delta R}{l_{\rm RM}}{-(\alpha+3)},&&l> L. 
\end{aligned}
\right.\label{sf2}
\end{align}
for a shallow spectrum ($-3<\alpha<-2$) and $L\sim\Delta R$. Therefore, 
the Kolmogorov scaling
with $\alpha=-11/3$ has the RM structure function $D_{\rm RM}(l)\propto l^{5/3}$ in the inertial range and $D_{\rm RM}(l)\sim\text{constant}$ beyond the inertial range.

We assume that the relative transverse velocity between the FRB source and the Faraday screen is $v_\perp$. 
Based on the definition of $D_{\rm RM}(l)$ in Eq.(\ref{SF}), the r.m.s. variation of RM during a time $t$ is
\be
|\delta{\rm RM}(t)|\sim \sqrt{D_{\rm RM}(v_\perp t)}.
\ee 
The largest RM amplitude contributed by the Faraday screen can be estimated as $|{\rm RM}|\sim|\delta {\rm RM}(l>L)|$, where $|\delta {\rm RM}(l>L)|\sim\sqrt{D_{\rm RM}(l> L)}$ is given by the last equations of Eq.(\ref{sf1}) and Eq.(\ref{sf2}). 
Thus, during time $t$, the relative RM variation can be estimated as
\begin{align}
\left|\frac{\delta {\rm RM}}{{\rm RM}}\right|\sim 
\left\{
\begin{aligned}
&\fraction{v_\perp t}{L}{-(\alpha+2)/2},&&v_\perp t< L,\\
&1,&&v_\perp t> L,
\end{aligned}
\right.\label{RMvariation}
\end{align}
for both steep and shallow spectra. We emphasize again that in the above equation $L\sim l_{\rm RM}$ is considered for a steep spectrum, and $L\sim\Delta R$ is considered for a shallow spectrum.

The measurements of the RM structure function of some FRB repeaters revealed that \citep{Mckinven22}
$D_{\rm RM}(t)\propto t^{0.2-0.4}$,
which implies that the power spectrum is 
$\alpha\sim-(2.2-2.4)$.
Therefore, the Faraday screens of these FRB repeaters have shallow spectra in the inertial ranges, which means that the variation is dominated by small-scale RM density fluctuations. 
Notice that since magnetic fluctuations arise from nonlinear turbulent dynamo \citep{Xu16} and turbulent compression causes the magnetic energy spectrum basically follows the turbulent energy spectrum,
the magnetic energy spectrum is usually steep ($\alpha<-3$).
Thus, the observed result implies that a shallow electron density spectrum is more likely to dominate the RM fluctuations for these particular FRB sources.
Physically, a shallow electron density spectrum naturally arises in supersonic turbulence, e.g., in star-forming regions \citep{Hennebelle12,Xu17}.

On the other hand, many FRB repeaters, e.g, FRB 121102, FRB 190520B, FRB 180916B, show large RM variations $|\delta{\rm RM}/{\rm RM}|\sim 1$ within a few months to a few years \citep{Michilli18,Hilmarsson21,Anna-Thomas22,Dai22,Mckinven22}, implying that the outer scale of the inertial range satisfies
\be
L\lesssim v_\perp t\simeq10^{-4}~{\rm pc}\fractionz{v_\perp}{100~{\rm km~s^{-1}}}\fractionz{t}{1~{\rm yr}}.\label{outerscale}
\ee

\subsection{Secular RM evolution}\label{secularRM}

A secular RM evolution may be attributed to the expansion of a magnetized shell or orbital motion of a binary system.
First, we consider the scenario of an expanding magnetized shell with the magnetic field configuration unchanged during a short term, which is applicable to young SNRs (see Section \ref{SNR}).
or companion flares in a binary system (see Section \ref{stellarflare}).
In a certain astrophysical environment, the electron density and magnetic field are usually related, e.g.
$B\propto n_e^{\gamma_B}$,
where $\gamma_B=1/2,2/3,1$ corresponds to an energy-equipartition plasma, a magnetic freezing plasma, or a shocked compressed plasma, respectively \citep[e.g.,][]{Yang22}. 
Due to the shell expansion, the electron density might decrease with the shell radius
$n_e\propto r^{-\gamma_n}$,
with $\gamma_n=0,2,3$ corresponding to a shock-compressed medium (upstream medium is assumed to be uniform), a wind medium, and free expansion, respectively. We assume that the time-dependent shell radius is $r\propto t^{\gamma_r}$ with $\gamma_r=1,2/5$ corresponding to free expansion and the Sedov-Taylor phase, respectively. Therefore, the RM evolution satisfies
\be
{\rm RM}\propto n_eBr\propto t^{\gamma_r(1-\gamma_n-\gamma_B\gamma_n)}.\label{shellRM}
\ee

Next, we are interested in the secular RM evolution caused by the orbital motion of a binary system, and the companion could be a stellar object (see Section \ref{stellarwind}) or a massive black hole (see Section \ref{BH}). Since the large-scale magnetic fields are contributed by the magnetized companion (i.e., the large-scale dipole field for a companion with weak wind, the magnetic field in the disk of Be stars, etc.), the RM variation would be periodic with the same period as the orbital period,
\be
P=2\pi\fraction{a^3}{GM_{\rm tot}}{1/2},\label{period}
\ee
where $a$ is the semi-major separation of the binary system, and $M_{\rm tot}$ is the binary total mass. A large RM variation of $|\delta {\rm RM}/{\rm RM}|\sim1$ should occur in the timescale of $\lesssim P$.
Such a scenario could be tested by long-term monitoring of RM variations for
particular FRB repeaters.

\section{Different astrophysical scenarios generating RM variations}\label{scenarios}

In this section, we will discuss a list of possible astrophysical processes that might cause RM variations of a particular FRB repeater. 

\subsection{RM variations contributed by a supernova remnant}\label{SNR}

Radio polarization observations of young SNRs suggest that 
magnetic fields in SNRs are largely disordered, with a small radial preponderance \citep[e.g.,][]{Dickel76,Milne87,Reynolds12}. There are two possible explanations to the radial preponderance \citep{Jun96,Blondin01,Zirakashvili08,Inoue13,West17}: 1. The Rayleigh-Taylor instability stretches the field lines preferentially along the radial direction; 2. Turbulence with a radially biased velocity dispersion may be induced. In older, larger SNRs, the field lines are often disordered but sometimes tangential. The tangential fields could be explained by the shock compression of the upstream medium \citep[e.g.,][]{Dickel76,Milne87,Reynolds12}.
In summary, polarization and imaging observations indicate that the magnetic fields in a SNR are turbulent and evolving.

\begin{figure}
    \centering
	\includegraphics[width = 1.0\linewidth, trim = 0 0 0 0, clip]{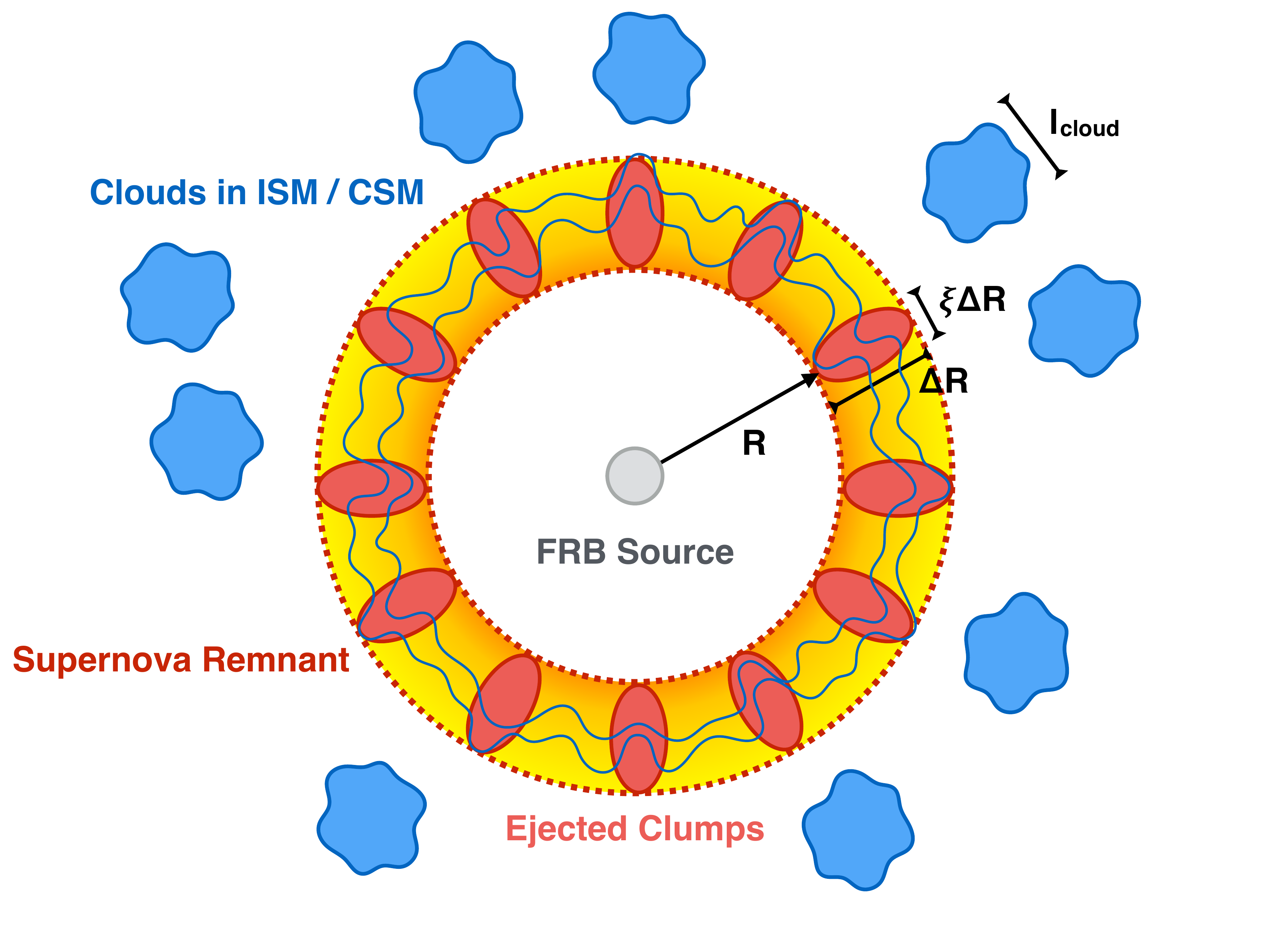}
    \caption{Schematic configuration of an FRB source in an SNR expanding in the interstellar/circumstellar medium with gas clouds. The SNR has radius $R$ and thickness $\Delta R$. The red regions correspond to ejected large-scale clumps in the SNR, which reflects the SNR local anisotropy. The blue regions correspond to the gas clouds with average scale $l_{\rm cloud}$ in the interstellar/circumstellar medium. The grey circle corresponds to the FRB source. The blue curves indicate magnetic field lines. The RM is contributed by the parallel component by the magnetic field fluctuations.}\label{figSNR} 
\end{figure}

If magnetic geometry along the line of sight keeps unchanged, the RM contribution from an SNR would show a long-term evolution due to the expansion of the SNR \citep[][see the discussion in Section \ref{FaradayScreen}]{Piro18,Zhao21b}, and the corresponding RM satisfies a power-law evolution given by Eq.(\ref{shellRM}). However, due to turbulence and relative motion between the FRB source and the SNR, the RM could show random variations in short term, especially for young SNRs. In the following discussion, we mainly focus on such a scenario.

When an SNR propagates in a highly inhomogeneous interstellar/circumstellar medium, as shown in Figure \ref{figSNR}, it generates a pair of shocks upon interacting with the dense gas clouds and induces turbulence in the magnetized medium \citep[e.g.,][]{Hu22}. 
Therefore, for the SNR scenario, one may take the largest outer scale of the turbulence in the SNR as 
\be
L=\min(\xi \Delta R,l_{\rm cloud}),\label{largestscale}
\ee 
where $\Delta R$ is the SNR thickness,  $\xi\Delta R$ is the transverse scale of the ejected large-scale clumps in the SNR (see Figure \ref{figSNR}), with the parameter $\xi$ describing the local anisotropy of the SNR (the smaller the value of $\xi$, the more significant the SNR local anisotropy), and $l_{\rm cloud}$ is the typical scale of the gas clouds \citep[e.g.,][]{Heiles03,Inoue09}.
The intensity map of the molecular emission shows that the density distribution within molecular clouds appears to have shallow spectra characterized by small-scale, large density. The smallest scale of clouds is 
\citep[see Figure 10 of][]{Hennebelle12}
\be
l_{\rm cloud}\lesssim 0.1~{\rm pc}.
\ee
Notice that the above upper limit of the characteristic size of density structures is due to the limited resolution of observations. 
According to Eq.(\ref{largestscale}), the time of the FRB source crossing the distance $L$ is
\begin{align}
t_L&\sim\frac{L}{v_\perp}=\frac{1}{v_\perp}\min(\xi \Delta R,l_{\rm cloud})\nonumber\\
&=1~{\rm yr}\fraction{v_\perp}{10^3~{\rm km~s^{-1}}}{-1}\fractionz{\min(\xi \Delta R,l_{\rm cloud})}{10^{-3}~{\rm pc}},\label{SNRvariation}
\end{align}
which corresponds to the typical timescale for $|\delta {\rm RM}/{\rm RM}|\sim1$.

We generally discuss a wide range of the possible transverse relative velocity as $v_\perp\sim(10-10^4)~{\rm km~s^{-1}}$. The lower limit of $v_\perp\sim10~{\rm km~s^{-1}}$ corresponds to the possible minimum intrinsic velocity of the neutron star FRB source \citep[e.g.,][]{Hansen97}, and the upper limit of $v_\perp\sim10^4~{\rm km~s^{-1}}$ corresponds to the initial velocity of an expanding SNR \citep[e.g.,][]{Yang17}. 

First, we consider that the scenario of $\xi\Delta R< l_{\rm cloud}$.
According to Eq.(\ref{RMvariation}) and Eq.(\ref{largestscale}), a relative RM variation $|\delta{\rm RM}/{\rm RM}|\sim 1$ during time $t$ implies $l\sim v_\perp t\sim\xi\Delta R\sim \xi\eta v_{\rm SNR}t_{\rm SNR}$, where $v_{\rm SNR}$ is the SNR expanding velocity, $t_{\rm SNR}$ is the SNR age, and the SNR thickness is $\Delta R=\eta R\sim\eta v_{\rm SNR}t_{\rm SNR}$, and $R$ is the SNR radius. Thus, the typical SNR age is given by
\be
t_{\rm SNR}\sim\fractionz{v_\perp}{\xi\eta v_{\rm SNR}}t.
\ee
For the SNR with a small radius of $R\sim\eta^{-1}\Delta R<(\xi\eta)^{-1} l_{\rm cloud}$ and for $0.01\lesssim\xi\eta\lesssim1$, the SNR might have a large expanding velocity of $v_{\rm SNR}\gtrsim10^3~{\rm km~s^{-1}}$ \citep[e.g.,][]{Yang17}. Considering that the intrinsic velocity of most neutrons stars might not exceed $\sim 10^3~{\rm km~s^{-1}}$ \citep[e.g.,][]{Hansen97}, one may have $v_\perp\lesssim v_{\rm SNR}$ in this scenario. 
The above result implies that the SNR has the typical age of $t_{\rm SNR}\lesssim100~{\rm yr}(\xi\eta/0.01)^{-1}(t/1~{\rm yr})$.  
Therefore, for some FRB repeaters showing $|\delta{\rm RM}/{\rm RM}|\sim 1$ within a few months to a few years, the SNR is required to be young with significant local anisotropy or a thin thickness, which would also show observable, secular evolution in both DM and RM \citep{Yang17,Piro18,Zhao21b}. 
For FRB repeaters with insignificant RM random variations, the typical timescale of the random variation would be much larger than the observing time of a few years. The corresponding SNRs would be older.

On the other hand, for the scenario of $\xi\Delta R>l_{\rm cloud}$, the SNR could be relatively older compared with the above scenario of $\xi\Delta R<l_{\rm cloud}$. According to Eq.(\ref{RMvariation}), Eq.(\ref{outerscale}), and Eq.(\ref{largestscale}), a significant relative RM variation of $|\delta{\rm RM}/{\rm RM}|\sim 1$ during observing time $t$ would give a constraint on the typical size of the gas clouds, i.e. $l_{\rm cloud}\lesssim10^{-4}~{\rm pc}(v_\perp/100~{\rm km~s^{-1}})(t/1~{\rm yr})$. 
Thus, for the FRB repeaters with insignificant RM random variations (e.g., the RM varies in a timescale of $t\gtrsim10^3~{\rm yr}$), gas clouds with moderate size ($l_{\rm cloud}\sim 0.1~{\rm pc}$) would satisfy the requirement.
However, if an FRB repeater shows significant RM variations within months to years, the cloud size would be constrained to be very small. Such small-scale density structures in SNRs are challenging to resolve observations due to the limited angular resolution \citep{Hennebelle12}.

\subsection{RM variations contributed by stellar winds from a massive/giant star companion}\label{stellarwind}

There is some evidence suggesting that an FRB source might be in a binary system: 1) FRB 180916B shows a periodic activity with a period of $16.35$ days \citep{CHIME20b}, which might correspond to an orbital period of a binary system as proposed by some authors \citep{Ioka20b,Dai20,Lyutikov20,Zhang20e,Li21d,wada21}; 2) FRB 20200120E was found to be associated with a globular cluster in the M81 galaxy \citep{Bhardwaj21,Kirsten22}, which contains many close binary systems; 3) PSR B1744-24A in a binary system displays a complex, magnetized environment with Faraday conversion and circularly polarized attenuation \citep{Li22}, which is similar to the properties of FRB 20201124A \citep{Xu21}; 4) RM variations consisting of a constant component and an irregular component have been observed both in FRB repeaters \citep{Xu21,Mckinven22} and pulsar binary systems \citep{Johnston96,Johnston05}. The latter requires an elliptical orbit of the pulsar in the binary system. A similar configuration may apply to FRB repeaters as well \citep[e.g.,][]{Li21d,Wang22}.

\begin{figure}
    \centering
	\includegraphics[width = 1.0\linewidth, trim = 0 0 0 0, clip]{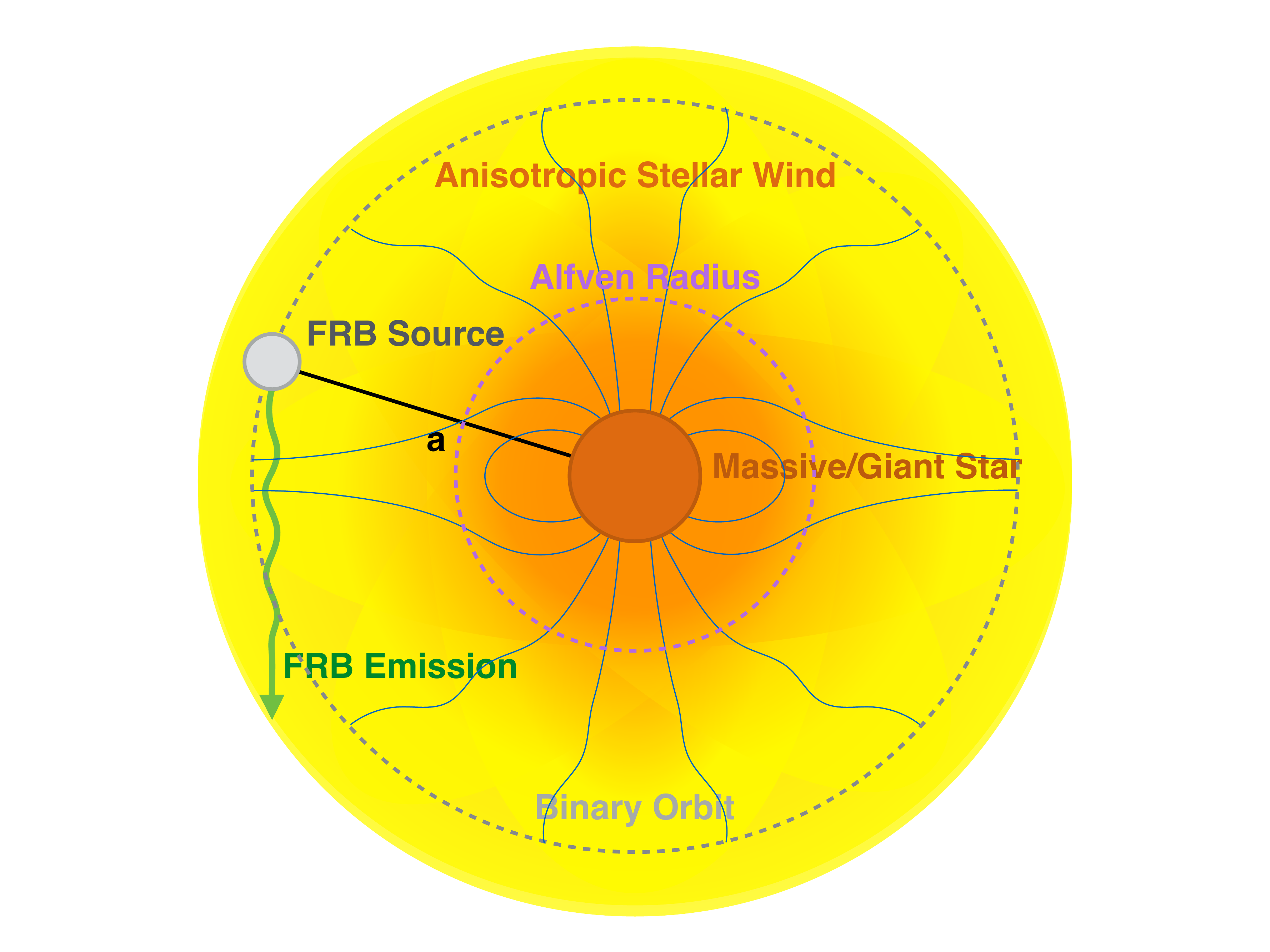}
    \caption{Schematic configuration of an FRB source in a binary system with the companion generating anisotropic stellar winds. The orange circle corresponds to the companion which might be a massive star or a giant star. The grey circle corresponds to the FRB source moving around the companion. The blue curves denote magnetic field lines.}\label{figstellarwind} 
\end{figure}

In this subsection, we consider the RM and its variation introduced by the stellar wind from a companion in a binary system, as shown in Figure \ref{figstellarwind}.
We assume that the companion star has a mass $M_c$, a radius $R_c$, and a mass loss rate of $\dot M$. The wind velocity can be calculated as the escape velocity, i.e. 
$v_w\sim(2GM_c/R_c)^{1/2}\sim620~{\rm km~s^{-1}}(M_c/M_\odot)^{1/2}(R_c/R_\odot)^{-1/2}$.
The electron density in the stellar wind at a distance $r$ from the star is given by
\begin{align}
n_w(r)&\simeq\frac{\dot M}{4\pi\mu_mm_pv_wr^2}\simeq1.1\times10^6~{\rm cm^{-3}}\nonumber\\
&\times\fraction{r}{1~{\rm AU}}{-2}\fractionz{\dot M}{10^{-8}M_\odot~{\rm yr^{-1}}}\fraction{v_w}{10^3~{\rm km~s^{-1}}}{-1},
\end{align}
where $\mu_m=1.2$ is the mean molecular weight for a solar composition, 
and the mass loss rate depends on the stellar type, e.g. $\dot M\sim10^{-7}-10^{-5}M_\odot~{\rm yr^{-1}}$ for O stars \citep{Puls96,Muijres12}; $\dot M\sim10^{-11}-10^{-8}M_\odot~{\rm yr^{-1}}$ for Be stars \citep{Snow81,Poe86}; $\dot M\sim10^{-14}-10^{-10}M_\odot~{\rm yr^{-1}}$ for solar-like stars \citep{Wood02}. We assume that the FRB emission region is close to the FRB source. Since the wind density decrease as $r^{-2}$, most of the local DM would be contributed by the wind at\footnote{Except for the scenario that the companion is in front of the FRB source along the line of sight, leading to a large DM during the eclipsing phase. A similar scenario has  been seen in PSR B1744-24A in the globular cluster Terzan 5 reported by \cite{Li22}. During the eclipsing phase, PSR B1744-24A shows a significant DM variation and depolarization caused by RM variation.} $r\sim a$, where $a$ is the binary separation. Thus, the DM contributed by the stellar wind is estimated as 
\begin{align}
{\rm DM}_w&\sim n_wa\simeq5.4~{\rm pc~cm^{-3}}\fraction{a}{1~{\rm AU}}{-1}\nonumber\\
&\times\fractionz{\dot M}{10^{-8}M_\odot~{\rm yr^{-1}}}\fraction{v_w}{10^3~{\rm km~s^{-1}}}{-1}.\label{windDM}
\end{align}
In order to estimate the RM contribution of the companion wind, we consider that the magnetic field strength at distance $r$ from the companion center satisfies
\begin{align}
B(r)\sim 
\left\{
\begin{aligned}
&B_c\fraction{r}{R_c}{-3},&&R_c<r<R_A,\\
&B_c\fraction{R_A}{R_c}{-3}\fraction{r}{R_A}{-\beta_B},&&R>R_A,
\end{aligned}
\right.
\end{align}
where $1\lesssim \beta_B\lesssim 2$. Typically, one has $\beta_B\simeq1$ for a toroidal field and $\beta_B\simeq2$ for a radial field. 
In the above equation, $B\propto r^{-3}$ corresponds to the dipole field near the companion surface and $B\propto r^{-\beta_B}$ corresponds to the magnetic field in the stellar wind outside the Alfv\'en radius $R_A$.
In the inner region $r<R_A$, the magnetic field pressure $P_B$ dominates, and the stellar wind moves along the field lines. In the outer region $r>R_A$, due to the large ram pressure of the stellar wind $P_w$, the magnetic field pressure is sub-dominant and would be carried by the wind. The Alfv\'en radius $R_A$ is defined by
\be
P_w=\rho_w(R_A)v_w^2\simeq\frac{\dot Mv_w}{4\pi R_A^2}\sim P_B=\frac{1}{8\pi}B_c(R_A)^2,
\ee
where $\rho_w=\dot M/(4\pi r^2 v_w)$ is the mass density of the stellar wind. 
According to the above equation, one has 
\begin{align}
R_A&\sim\fraction{B_c^2R_c^6}{2\dot Mv_w}{1/4}\simeq0.08R_\odot\fraction{B_c}{1~{\rm G}}{1/2}\fraction{R_c}{1R_\odot}{3/2}\nonumber\\
&\times\fraction{\dot M}{10^{-8}M_\odot~{\rm yr^{-1}}}{-1/4}\fraction{v_w}{10^3~{\rm km~s^{-1}}}{-1/4},\label{alfven}
\end{align}
suggesting that $R_A$ is inside the star for our typical parameters, so that 
the magnetic field is dominated by the stellar wind at $r>R_c$.
In other words, the stellar wind would carry a relatively strong magnetic field extending to large distances.
The RM contributed by the companion wind is approximately
\begin{align}
{\rm RM}_w&\sim\frac{e^3Bn_wa}{2\pi m_e^2c^4} \simeq2\times10^4~{\rm rad~m^{-2}}\fractionz{B_c}{1~{\rm G}}\fractionz{R_c}{1R_\odot}\nonumber\\
&\times\fractionz{\dot M}{10^{-8}M_\odot~{\rm yr^{-1}}}\fraction{a}{1~{\rm AU}}{-2}\fraction{v_w}{10^3~{\rm km~s^{-1}}}{-1}.\label{RMw1}
\end{align}
for a toroidal field with $\beta_B\simeq1$, and 
\begin{align}
{\rm RM}_w&\simeq94~{\rm rad~m^{-2}}\fractionz{B_c}{1~{\rm G}}\fractionz{R_c}{1R_\odot}\nonumber\\
&\times\fractionz{\dot M}{10^{-8}M_\odot~{\rm yr^{-1}}}\fraction{a}{1~{\rm AU}}{-3}\fraction{v_w}{10^3~{\rm km~s^{-1}}}{-1}.\label{RMw2}
\end{align}
for a radial field with $\beta_B\simeq2$.
Based on the above results, the environment of FRB repeaters with large RMs might correspond to the stellar wind of a massive star or a giant star with a toroidal magnetic field configuration.

When an FRB repeater is in a binary system, both the orbital motion of the binary system and the dynamic evolution of the companion wind could cause RM variation. 
As discussed in Section \ref{FaradayScreen}, a magnetized companion usually has a large-scale field, and the orbital motion of the radio source in such a large-scale field would cause RM variation \citep{Wang22,Li22}. 
Such a scenario has been observed in a binary system. For example, the RM of PSR B1259-63 reaches a few times $10^3~{\rm rad~m^{-2}}$ and significantly reverses sign around the periastron \citep{Johnston96,Johnston05}.
Consider that the total mass of the binary system is $M_{\rm tot}$. The orbital period is given by Eq.(\ref{period}), i.e.
\be
P
\simeq0.3~{\rm yr}\fraction{a}{1~{\rm AU}}{3/2}\fraction{M_{\rm tot}}{10~M_\odot}{-1/2}.
\ee
In this scenario, RM variation should have the same period as the orbital period, which could be tested by long-term observation. 

Turbulence in stellar winds could be caused by the anisotropic distribution or the episodic outflow of the stellar wind, which may smear out the apparent RM periodic evolution.
For turbulence of the companion wind at a distance $r\sim a$, the typical outer scale of turbulence might be estimated by
\be
L\sim \min(\theta_w a, v_w\Delta t_w),
\ee
where $\theta_w$ is the typical anisotropic distribution angle of the stellar wind, and $\Delta t_w$ is the typical timescale of the wind outflow variation. The episodic wind variation is usually caused by stellar flares, which will be further discussed in Section \ref{stellarflare}. Here we are mainly interested in the case of persistent wind with $\theta_w a\ll v_w\Delta t_w$, leading to
\be
L\sim \theta_w a\simeq1~{\rm AU}\fractionz{\theta_w}{1~{\rm rad}}\fractionz{a}{1~{\rm AU}},
\ee
The Keplerian velocity of the FRB source around the companion is
\be
v=\fraction{GM_{\rm tot}}{a}{1/2}\simeq94~{\rm km~s^{-1}}\fraction{M_{\rm tot}}{10M_\odot}{1/2}\fraction{a}{1~{\rm AU}}{-1/2}.
\ee
Therefore, the time of the FRB source crossing the outer scale $L$ is 
\be
t_L\sim\frac{L}{v}\simeq18~{\rm day}\fractionz{\theta_w}{1~{\rm rad}}\fraction{M_{\rm tot}}{10M_\odot}{-1/2}\fraction{a}{1~{\rm AU}}{3/2}.\label{windvariation}
\ee
According to Eq.(\ref{sf1}), Eq.(\ref{sf2}) and Eq.(\ref{RMvariation}), one has $|\delta {\rm RM}/{\rm RM}|\propto t^{-(\alpha+2)/2}$ and $D_{\rm RM}(t)\propto t^{-(\alpha+2)}$ for $t\lesssim t_L$; and $|\delta {\rm RM}/{\rm RM}|\sim 1$ and $D_{\rm RM}(t)\sim\text{constant}$ for $t\gtrsim t_L$.

In particular, if the binary orbit is elliptical, a larger RM variation would occur near the periastron due to a stronger magnetic field and a higher electron density and would keep almost constant far away from the periastron, as was observed in PSR B1259-63 \citep{Johnston96,Johnston05}. Some FRB repeaters, e.g., FRB 20201124A, also exhibited the similar behaviors \citep{Xu21,Wang22}, although the periodic evolution of RM fluctuations has not been detected so far. 
It is worth noting that for this scenario, the periodic evolution of RM variations would be easier to achieve for an elliptical orbit compared with a circular orbit, 
because the possible significant fluctuations of electron density and magnetic fields in the latter case might smear out the clean periodic signature in RM variation.

\subsection{RM variations contributed by stellar flares from a low-mass star companion}\label{stellarflare}

Stellar flares are usually defined as catastrophic releases of magnetic energy leading to particle acceleration and electromagnetic radiation accompanied by coronal mass ejections (CMEs) \citep[e.g.,][]{Haisch91}. Frequent flaring occurs on stars with an outer convection zone, and the timescale of energetic fares is longer than that of less energetic flares \citep[e.g.,][]{Pettersen89}. 
Over short timescales of minutes to a few hours, they emit energy 
ranging from $10^{23}~{\rm erg}$ (nanoflares; e.g., \cite{Parnell00}) to 
$10^{31}-10^{38}~{\rm erg}$ (superflare; e.g., \cite{Shibayama13,Gunther20}).
In particular, for low-mass stars, due to  strong convection near their surfaces, flares and CMEs are usually frequent. In the following discussion, we will mainly focus on stellar flares/CMEs from low-mass stars.

\begin{figure}
    \centering
	\includegraphics[width = 1.0\linewidth, trim = 0 0 0 0, clip]{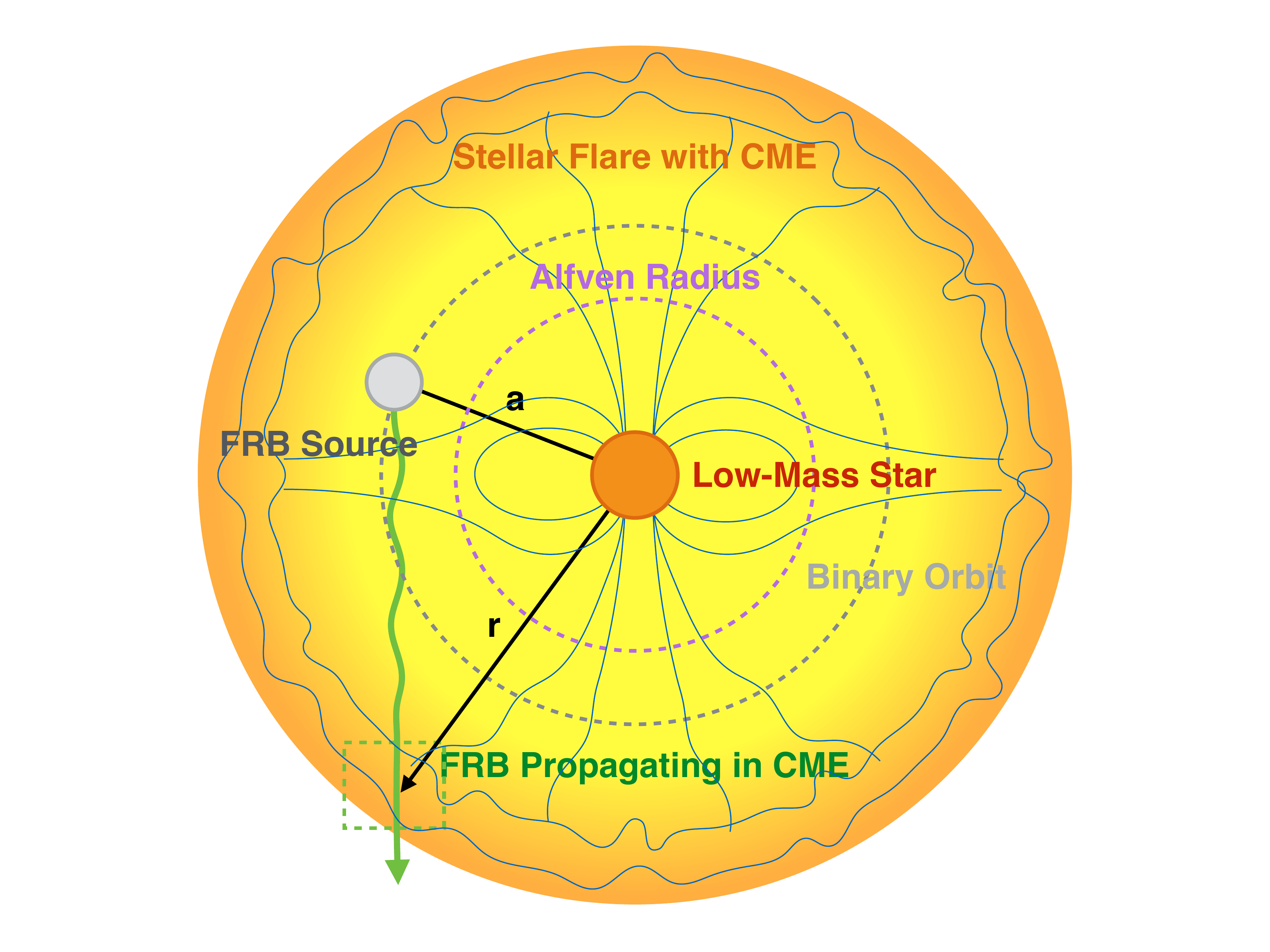}
    \caption{Schematic configuration of an FRB source in a binary system with the companion producing CMEs associated with flares. The orange shell corresponds to the CME where the FRB crosses it at the distance $r$ from the center of the companion star.
    The orange circle corresponds to the companion that might be a low-mass star. The grey circle corresponds to the FRB source moving around the companion. The blue curves denote magnetic field lines.}\label{figstellarflare} 
\end{figure}

To estimate the impact of stellar flares/CMEs, we apply the empirical relationship between flare energy in the X-ray band, $E_X$, and the CME mass, $M_{\rm CME}$, found by \cite{Aarnio12}, i.e. $\log M=0.63\log E_X-2.57$. We assume that the ratio between the stellar flare X-ray energy $E_X$ and the CME kinetic energy $E_{\rm CME}$ is $\epsilon$, i.e. $E_X=\epsilon_X E_{\rm CME}$, and adopt a typical value $\epsilon_X\sim0.01$ 
considering that the energy emitted bolometrically is typically larger than the X-ray energy by a factor of 100 for the same flare strength \citep{Osten15,Gunther20}. Therefore, the CME mass is 
\be
M_{\rm CME}\simeq 2.1\times10^{16}~{\rm g}\fraction{E_{\rm CME}}{10^{32}~{\rm erg}}{0.63},
\ee
and the CME velocity is estimated as
\be
v_{\rm CME}\simeq\fraction{2E_{\rm CME}}{M_{\rm CME}}{1/2}
\simeq980~{\rm km~s^{-1}}\fraction{E_{\rm CME}}{10^{32}~{\rm erg}}{0.185}.
\ee
We define $t_{1/2}$ as the decay time from the peak luminosity to half of that level for a stellar flare. The U-filter flare data show that $t_{1/2}$ depends on the U-band flare energy, i.e. $\log t_{1/2}=0.3\log E_U-7.5$ in cgs unit \citep{Pettersen89}. We assume $E_U=\epsilon_U E_{\rm CME}$ with the efficiency taken as $\epsilon_U\sim0.1$ for U band \citep[e.g.,][]{Osten15}. The flare duration $\Delta t$ is then estimated as 
\be
\Delta t\sim t_{1/2}\simeq63~{\rm s}\fraction{E_{\rm CME}}{10^{32}~{\rm erg}}{0.3}.
\ee
Therefore, the electron density of the CME at the distance $r$ from the center of the companion star is
\begin{align}
n_{\rm CME}(r)&\simeq\frac{M_{\rm CME}}{4\pi\mu_mm_pv_{\rm CME}r^2\Delta t}\nonumber\\
&\simeq600~{\rm cm^{-3}}\fraction{E_{\rm CME}}{10^{32}~{\rm erg}}{0.145}\fraction{r}{1~{\rm AU}}{-2}.
\end{align}
Different from the scenario of stellar winds, the contributions of DM and RM from the stellar flare mainly depend on the time difference between the stellar flare and FRBs. We consider that an FRB encounters the stellar flare at the distance $r$ from the companion star, as shown in Figure \ref{figstellarflare}. The DM contribution by the flare is approximately
\be
{\rm DM}_{\rm CME}\sim n_{\rm CME}r\simeq2.9\times10^{-3}~{\rm pc~cm^{-3}}\fraction{E_{\rm CME}}{10^{32}~{\rm erg}}{0.145}\fraction{r}{1~{\rm AU}}{-1}.\label{flareDM}
\ee
Similar to the discussion of stellar winds in Section \ref{stellarwind}, the Alfv\'en radius of the stellar flare is given by Eq.(\ref{alfven}), i.e.
\begin{align}
R_A&\sim\fraction{B_c^2R_c^6\Delta t}{2M_{\rm CME}v_{\rm CME}}{1/4}\nonumber\\
&\simeq0.5R_\odot\fraction{B_c}{1~{\rm G}}{1/2}\fraction{R_c}{1R_\odot}{3/2}\fraction{E_{\rm CME}}{10^{32}~{\rm erg}}{-0.13}.
\end{align}
Because $r\gg R_c\gtrsim R_A$, the magnetic field in the stellar flare also satisfies $B(r)\propto r^{-\beta_B}$, with $\beta_B$ ranging from 1 to 2. Therefore, the RM contribution by a flare is approximately
\begin{align}
{\rm RM}_{\rm CME}&\sim \frac{e^3}{2\pi m_e^2c^4} Bn_{\rm CME}r\simeq11~{\rm rad~m^{-2}}\nonumber\\
&\times\fractionz{B_c}{1~{\rm G}}\fractionz{R_c}{1R_\odot}\fraction{E_{\rm CME}}{10^{32}~{\rm erg}}{0.145}\fraction{r}{1~{\rm AU}}{-2}.\label{RMflare1}
\end{align}
for a toroidal field with $\beta_B\simeq1$, or 
\be
{\rm RM}_{\rm CME}\simeq 0.05~{\rm rad~m^{-2}}\fractionz{B_c}{1~{\rm G}}\fractionz{R_c}{1R_\odot}\fraction{E_{\rm CME}}{10^{32}~{\rm erg}}{0.145}\fraction{r}{1~{\rm AU}}{-3}.\label{RMflare2}
\ee
for a radial field with $\beta_B\simeq2$. Notice that the above typical values of RMs are much smaller than those of stellar winds given by Eq.(\ref{RMw1}) and Eq.(\ref{RMw2}) for the same parameters. This is because the absolute mass loss rates of stellar flares/CMEs from low-mass stars are much smaller than those of stellar winds from massive/giant stars.

In particular, for a low-mass star with a surface magnetic field $B_c\sim10^3~{\rm G}$ and a radius of $R_c\sim0.1R_\odot$ (due to strong convection near the surface of a low-mass star, its surface magnetic fields are usually stronger than those of a massive star, also see \cite{Kochukhov21}), the RM contributed by its flare at $r\sim1~{\rm AU}$ could reach $\sim10^3~{\rm rad~m^{-2}}$ for a toroidal field. 
Therefore, if the large RMs of $\gtrsim10^3~{\rm rad~m^{-2}}$ are contributed by the companion flare, the separation of such a binary system is required to be small, with $a\lesssim1~{\rm AU}$, and each radio burst is required to be emitted just when the CME sweeps the FRB source. These requirements are likely fine-tuning.
Considering that the burst rate of radio bursts of some FRB repeaters ($\gtrsim 100~\text{bursts}~{\rm day^{-1}source^{-1}}$ for FRB 121102, \cite{Lid21}) is much larger than the rate of stellar flares ($\lesssim10~\text{flares}~{\rm day^{-1}source^{-1}}$ e.g., \cite{Osten15,Davenport16}) and that FRB emission and stellar flaring in a binary system should be physically independent,  the time difference between FRBs and flares should be randomly distributed.
We assume that two radio bursts (Burst A and Burst B) are emitted with a ten-day time delay in a binary system with a separation of $a\sim1~{\rm AU}$ and that the companion star has $B_c\sim10^3~{\rm G}$ and $R_c\sim0.1R_\odot$. Burst A is emitted when the CME just sweeps the FRB source, leading to ${\rm RM}\sim10^3~{\rm rad~m^{-2}}$. 
Assuming that the CME velocity is $v\sim1000~{\rm km~s^{-1}}$, the CME would be at $r\sim6.8~{\rm AU}$ when Burst B is crossing the CME. The RM of Burst B is about ${\rm RM}\sim22~{\rm rad~m^{-2}}$. In summary, the RM varies from $\sim 10^3~{\rm rad~m^{-2}}$ to $\sim 10~{\rm rad~m^{-2}}$ during ten days. Such an extreme RM variation has not been observed in any FRB repeaters, which is inconsistent with the current observations unless flares are more frequent so that Burst B would encounter another newly ejected CME when it is observed. 

\subsection{RM variations contributed by pulsar winds, pulsar wind nebulae, or magnetar flares}\label{pair}

\begin{figure}
    \centering
	\includegraphics[width = 1.0\linewidth, trim = 0 150 0 100, clip]{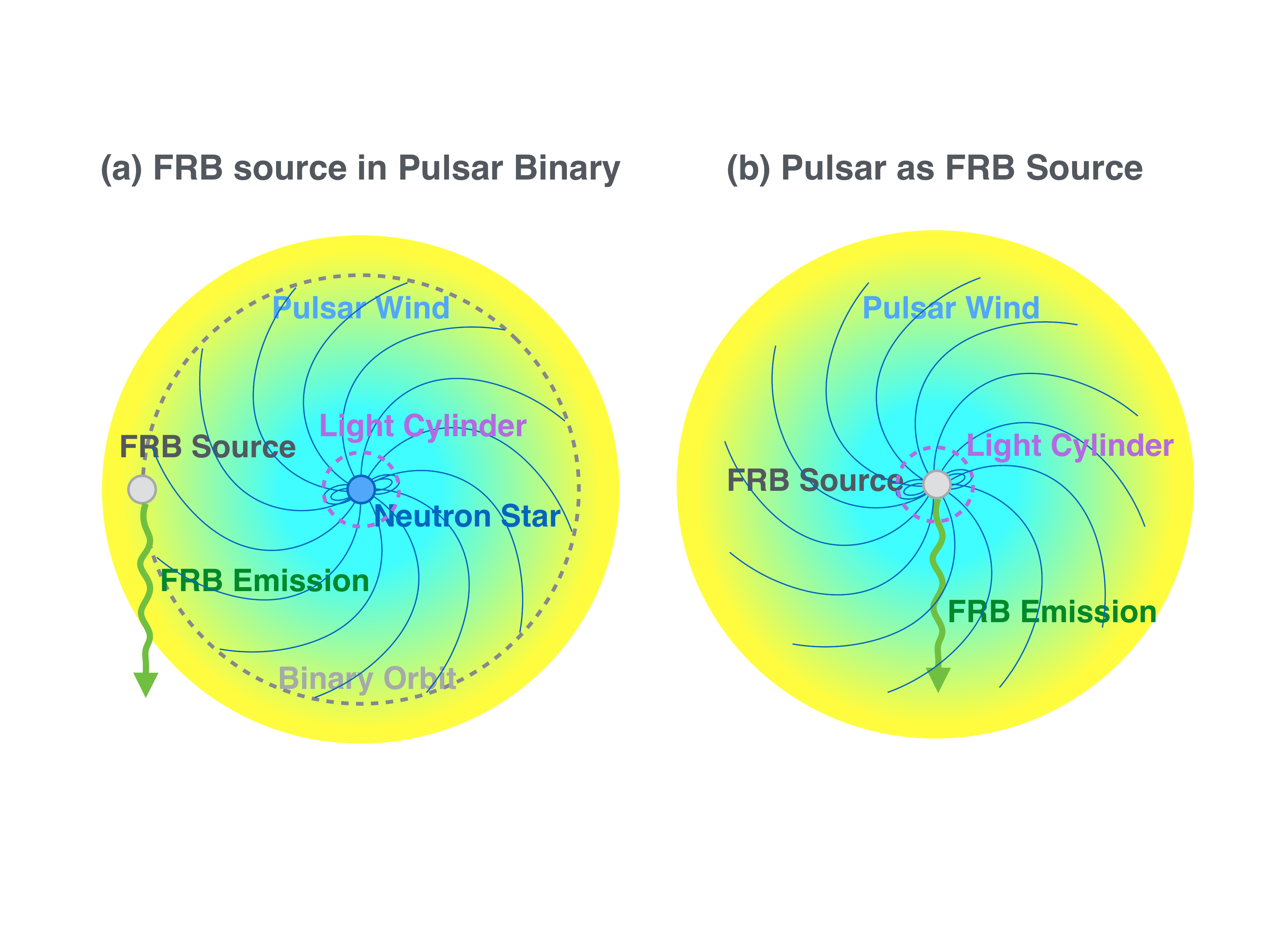}
    \caption{Schematic configuration of an FRB source in a pulsar wind. There are two possibilities: (a) the FRB source is in a binary system with a pulsar as its companion. (b) the pulsar wind is produced by the FRB source itself. The grey circle corresponds to the FRB source, and the blue circle corresponds to the companion pulsar. The blue curves denote magnetic field lines.}\label{figpulsarwind} 
\end{figure}

In this subsection, we consider the RM contribution from a pair plasma, including a pulsar wind, a pulsar wind nebula, or a magnetar flare. A pulsar wind could be produced by a neutron star as the companion of the FRB source in a binary system (see panel (a) of Figure \ref{figpulsarwind}) or by the FRB source itself (see panel (b) of Figure \ref{figpulsarwind}). 
In order to place a strong constraint, we first assume that the FRB source is at the center of the pulsar wind in the following discussion (Figure \ref{figpulsarwind}b). If the FRB source significantly deviates from the center of the pulsar wind (Figure \ref{figpulsarwind}a), the corresponding RM would be smaller. 
In general, since the pulsar wind is composed of relativistic electron-positron pairs, its RM contribution would be very small as proved below.

For a neutron star with radius $R$, dipole magnetic field at the pole $B_p$, and angular velocity $\Omega$, the spin-down power of the neutron star is 
\be
L_{\rm sd} \simeq \frac{B_p^2R^6\Omega^4}
{6c^3}
\simeq9.6\times10^{36}~{\rm erg~s^{-1}}\fraction{B_{p}}{10^{13}~{\rm G}}{2}\fraction{P}{0.1~{\rm s}}{-4}.
\ee
The magnetic field is nearly dipolar inside the light cylinder $R_{\rm LC}=c/\Omega$ but becomes toroidal in the pulsar wind. Thus, the field strength at $r>R_{\rm LC}$ is given by
\begin{align}
B(r)&=\frac{B_p}{2}\left(\frac{R}{R_{\rm LC}}\right)^{3}\left(\frac{R_{\rm LC}}{r}\right)\nonumber\\
&=2.2~{\rm G}\fractionz{B_p}{10^{13}~{\rm G}}\fraction{P}{0.1~{\rm s}}{-2}\fraction{r}{10^{13}~{\rm cm}}{-1},\label{B}
\end{align}
where $B_p/2$ is the mean surface magnetic field strength.
The Goldreich-Julian particle ejection rate from the polar cap may be estimated by
\be
\dot N_{\rm GJ}=2cA_{\rm cap}n_{\rm GJ}=2.7\times10^{33}~{\rm s^{-1}}\fractionz{B_p}{10^{13}~{\rm G}}\fraction{P}{0.1~{\rm s}}{-2},
\ee
where $n_{\rm GJ}=B_p/Pec$ is the Goldreich-Julian density at the neutron star pole \citep{Goldreich69}, and $A_{\rm cap}\simeq\pi R^3/R_{\rm LC}$ is the area of the polar cap for $R_{\rm LC}\gg R$. 
For a pair multiplicity $\mathcal{M}$, the electron/positron number density at distance $r$ is 
\begin{align}
n_{e}(r)&=\frac{\mathcal{M}\dot N_{\rm GJ}}{4\pi c r^2}=0.07~{\rm cm^{-3}}\fractionz{\mathcal{M}}{10^3}\nonumber\\
&\times\fractionz{B_p}{10^{13}~{\rm G}}\fraction{P}{0.1~{\rm s}}{-2}\fraction{r}{10^{13}~{\rm cm}}{-2}.\label{n}
\end{align}
We consider that the comoving Lorentz factor of the pulsar wind is $\gamma$, and the pair plasma is hot with thermal Lorentz factor $\gamma_{\rm th}$ in the comoving frame. 
For the pair plasma, due to the symmetry of positive and negative charges, its RM would be suppressed by the multiplicity $\mathcal{M}$. In other words, only the net charges contribute to Faraday rotation. 
Meanwhile, in the comoving frame, the RM contribution from relativistic hot electrons is suppressed by a factor of $\gamma_{\rm th}^2$ due to the relativistic mass $m_e\rightarrow\gamma_{\rm th} m_e$ \citep{Quataert00}.
Therefore, the RM contributed by the pulsar wind is (see Appendix \ref{RMpair})
\be
{\rm RM}=\frac{e^3}{\pi m_e^2c^4}\frac{1}{\gamma_{\rm th}^2\mathcal{M}}\int_{r_c}^dn_eB_\parallel ds,\label{RMw}
\ee
Notice that the above equation does not directly involve the wind Lorentz factor $\gamma$ (see Appendix \ref{RMpair} for details). 
In Eq.(\ref{RMw}), $d$ corresponds to the radius of a pulsar wind nebula, and $r_c$ is the critical radius where the electron cyclotron frequency is equal to the wave frequency in the comoving frame, leading to 
\be
B(r_c)\sim \frac{2\pi m_ec}{e}\left(\frac{\nu}{\gamma}\right)\simeq3.6~{\rm G}\fraction{\gamma}{100}{-1}\fractionz{\nu}{1~{\rm GHz}}.\label{Brc}
\ee
Different from Eq.(\ref{Bc}), here a factor of $1/\gamma$ is involved due to the Doppler effect of relativistic motion of the pulsar wind and the unchanged parallel field in different frames (see Appendix \ref{RMpair}).
In the region with $r<r_c$, the requirement of $\psi\propto\lambda^2$ for the RM measurement could not be satisfied according to the wave dispersion relation (see Appendix \ref{RMpair}).
According to the equations of $B(r)$ and $n_e(r)$ (Eq.(\ref{B}) and Eq.(\ref{n})), one has
\be
\frac{B}{n_e}\simeq30~{\rm G~cm^3}\fraction{\mathcal{M}}{10^3}{-1}\fractionz{r}{10^{13}~{\rm cm}},\label{Bn}
\ee
which is independent of the surface magnetic field and period of the neutron star.
Due to $n_e\propto r^{-2}$ and $B\propto r^{-1}$, most RM is contributed at $r_c$. Thus, according to Eq.(\ref{RMw}), Eq.(\ref{Brc}), and Eq.(\ref{Bn}), the RM contributed by a pulsar wind is estimated by
\be
{\rm RM}\simeq2.3\times10^{-3}~{\rm rad~m^{-2}}\gamma_{\rm th}^{-2}\fraction{\gamma}{100}{-2}\fraction{\nu}{1~{\rm GHz}}{2},\label{RMp}
\ee
which is very small. Note that in the above equation, the frequency-dependent RM is due to the integral lower limit $r_c$. For an FRB with a finite bandwidth between $(\nu_{\min},\nu_{\max})$, the RM measurement based on the condition $\psi={\rm RM}\lambda^2$ implies that $\nu$ in Eq.(\ref{RMp}) should be replaced by the observed minimum frequency $\nu_{\rm min}$.
Notice that RM mainly depends on pulsar wind properties considering that the FRB emission region is usually at $<r_c$, while DM more sensitively depends on the FRB emission region and pair production details in the magnetosphere.

In conclusion, the RM and its variation contributed by a pulsar wind is very small, which is independent of the surface magnetic field, period and multiplicity of the neutron star. 
Note that the above discussion assumes that the parallel component of the magnetic field is of the order of the total field, $B_\parallel\sim B$. For the pulsar wind scenario with the field almost perpendicular to the wind velocity \citep[e.g.,][]{Becker09}, the parallel component $B_\parallel$ would be much smaller, leading to an even smaller RM contribution. 

We are also interested in the scenarios of pulsar wind nebulae and magnetar flares. A pulsar wind nebula is produced by the interaction between the pulsar wind and the SNR/interstellar medium. In this process, the kinetic energy of the pulsar wind is transferred to thermal energy, i.e., $\gamma\rightarrow\gamma_{\rm th}$. Meanwhile, more pairs could be generated via magnetic reconnection, but the number of net charges keeps unchanged. Because $n_e\propto r^{-2}$ and $B\propto r^{-1}$, at the pulsar wind nebula radius that is much larger than $r_c$, the RM contribution would be much smaller. 

For magnetar flares, since a part of flare energy is transferred to relativistic pairs \citep{Thompson95}, their RM contribution is also expected to be very small for the same reason as in the pulsar winds and pulsar wind nebulae.
The estimated RM contributions from pulsar winds (nebulae) and magnetar flares are also consistent with the observations of most Galactic radio pulsars and radio-loud magnetars that have relatively small RMs mainly contributed from the interstellar medium. 
In particular, FRB 200428 was produced during the active phase of the magnetar SGR J1935+2152 \citep{Bochenek20,CHIME20}, which was associated with a hard X-ray burst \citep{Mereghetti20,Li20,Ridnaia20,Tavani20}. However, its RM is almost consistent with the value during its radio pulsar phase (Zhu et al. 2022, submitted). This implies that a magnetar flare cannot contribute significantly to RM variations. 

\subsection{RM variations contributed by magnetized outflows from a massive black hole}\label{BH}

\begin{figure}
    \centering
	\includegraphics[width = 1.0\linewidth, trim = 0 0 0 0, clip]{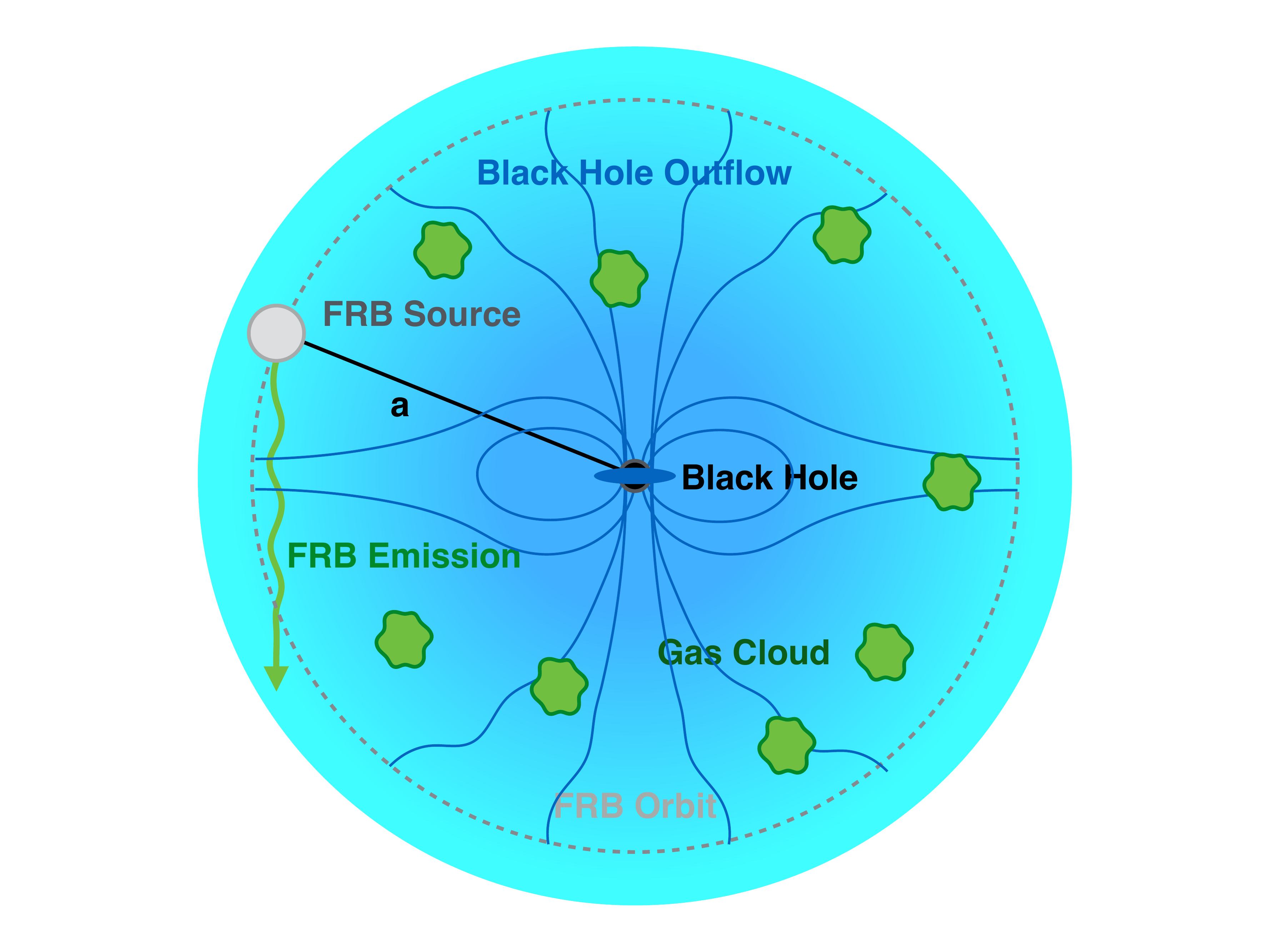}
    \caption{Schematic configuration of an FRB source in the vicinity of a massive black hole generating anisotropic outflow. The central black circle with a blue bar corresponds to the massive black hole with an accretion disk. The grey circle corresponds to the FRB source moving orbiting the massive black hole. The green regions correspond to the gas clouds near the massive black hole. The blue curves correspond to the magnetic field lines.}\label{figblackhole} 
\end{figure}

The extremely large RMs with ${\rm RM}\gtrsim 10^4~{\rm rad~m^{-2}}$ have been observed in the vicinities of massive black holes \citep{Bower03,Marrone07,Eatough13}. For example, the radio-loud magnetar PSR J1745-2900, which resides just $0.12~{\rm pc}$ from Sgr ${\rm A}^\ast$ \citep{Eatough13}, shows a large but relatively stable DM of $1800~{\rm pc~cm^{-3}}$ (consistent with a source located within $<10~{\rm pc}$ of the Galactic center, in the framework of the NE 2001 free electron density model of the Galaxy \citep{Cordes02}) and RM of $10^4-10^5~{\rm rad~m^{-2}}$ with an RM variability of $\sim 3500~{\rm rad~m^{-2}}$ \citep{Desvignes18}. Thus, 
it has been suggested that the large RMs observed in some FRB repeaters might be a result that the source is located in the vicinity of a massive black hole\footnote{Similar to massive black holes, for stellar-mass accreting black holes, \citet{Sridhar22} proposed that the highly compact luminous baryon-rich ``hyper-nebulae'' with the large mass loss from the disk winds and the polar jet might cause the significant RM variation for some FRB repeaters.} \citep{Michilli18,Zhang18,Anna-Thomas22,Dai22}, as shown in Figure \ref{figblackhole}.

Since the wind from a massive black hole is attributed to its accretion, the mass loss rate of the massive black hole can be normalized to the Eddington accretion rate $\dot M_{\rm Edd}$ with a dimensionless parameter $f$,
\begin{align}
\dot M&=f\dot M_{\rm Edd}=\frac{4\pi Gm_p}{\epsilon_{\rm BH}\sigma_Tc}fM_{\rm BH}\nonumber\\
&\simeq2.2\times10^{-3}M_\odot~{\rm yr^{-1}}f\fractionz{M_{\rm BH}}{10^5M_\odot},
\end{align}
where $M_{\rm BH}$ is the black hole mass, and $\epsilon_{\rm BH}\sim0.1$ is the radiative efficiency of a black hole accretion disk. Thus, the electron density at distance $r$ from the massive black hole is
\begin{align}
n_e(r)&\simeq\frac{\dot M}{4\pi\mu_mm_pvr^2}\nonumber\\
&\simeq2\times10^3~{\rm cm^{-3}}f\fractionz{M_{\rm BH}}{10^5M_\odot}\fraction{v}{0.1c}{-1}\fraction{r}{10^{-2}~{\rm pc}}{-2}.
\end{align}
Similar to the discussion on stellar winds (see Section \ref{stellarwind}), the DM contribution is 
\be
{\rm DM}\sim n_ea
\simeq20~{\rm cm^{-3}}f\fractionz{M_{\rm BH}}{10^5M_\odot}\fraction{v}{0.1c}{-1}\fraction{a}{10^{-2}~{\rm pc}}{-1},\label{BHDM}
\ee
where $a$ is the separation between the FRB source and the massive black hole,
and the RM contribution is
\begin{align}
{\rm RM}&\sim \frac{e^3}{2\pi m_e^2c^4} B_rn_ea\simeq1.6\times10^4~{\rm rad~m^{-2}}f\nonumber\\
&\times\fractionz{M_{\rm BH}}{10^5M_\odot}\fraction{v}{0.1c}{-1}\fractionz{B_r}{1~{\rm mG}}\fraction{a}{10^{-2}~{\rm pc}}{-1},\label{RMBH}
\end{align}
where $B_r$ is the field strength at $r\sim a$ from the center of the massive black hole.

When an FRB source is moving near a massive black hole with a Keplerian velocity of
\be
v\simeq\fraction{GM_{\rm BH}}{a}{1/2}\simeq210~{\rm km~s^{-1}}\fraction{M_{\rm BH}}{10^5M_\odot}{1/2}\fraction{a}{10^{-2}~{\rm pc}}{-1/2},
\ee
the RM variation is accounted for by the changes of magnetic field or electron density due to orbital motion or the inhomogeneous wind medium similar to the scenarios of binary systems. 
Different from the stellar wind scenario, the outflow from a massive black hole might interact with the gas clouds in the vicinity of the massive black hole.
For turbulence in the outflow at the distance $r\sim a$, the typical turbulence outer scale may be estimated as
\be
L\sim \min(\theta_w a, l_{\rm cloud}),
\ee
where $\theta_{\rm out}$ is the typical anisotropic distribution angle of the outflow from the massive black hole, and $l_{\rm cloud}$ is the typical size of the clouds for the turbulence induced by the interaction of the outflow with the circumnuclear clouds. If anisotropy of the outflow is significant, $\theta_w a<l_{\rm cloud}$, one has
\be
L\sim \theta_w a\simeq10^{-2}~{\rm pc}\fractionz{\theta_{\rm out}}{1~{\rm rad}}\fractionz{a}{10^{-2}~{\rm pc}},
\ee
and the timescale of the FRB source crossing the outer scale $L$ is 
\be
t_L\sim\frac{L}{v}\simeq47~{\rm yr}\fractionz{\theta_{\rm out}}{1~{\rm rad}}\fraction{M_{\rm BH}}{10^5M_\odot}{-1/2}\fraction{a}{10^{-2}~{\rm pc}}{3/2}.\label{variationBH1}
\ee
If the typical cloud size is small with $l_{\rm cloud}<\theta_w a$, one has
\be
L\sim l_{\rm cloud}\simeq10^{-2}~{\rm pc}\fractionz{l_{\rm cloud}}{10^{-2}~{\rm pc}},
\ee
and the time of the FRB source crossing the outer scale $L$ is 
\be
t_L\sim47~{\rm yr}\fractionz{l_{\rm cloud}}{10^{-2}~{\rm pc}}\fraction{M_{\rm BH}}{10^5M_\odot}{-1/2}\fraction{a}{10^{-2}~{\rm pc}}{1/2}.\label{variationBH2}
\ee
According to Eq.(\ref{sf1}), Eq.(\ref{sf2}) and Eq.(\ref{RMvariation}), one has $|\delta {\rm RM}/{\rm RM}|\propto t^{-(\alpha+2)/2}$ and $D_{\rm RM}(t)\propto t^{-(\alpha+2)}$ for $t\lesssim t_L$; and $|\delta {\rm RM}/{\rm RM}|\sim 1$ and $D_{\rm RM}(t)\sim\text{constant}$ for $t\gtrsim t_L$. At last, if the orbit of the FRB source is elliptical, a large RM variation would occur near the periastron and keep almost constant far away from the periastron, and a significant periodic evolution of RM variation could be tested through long-term monitoring of the FRB sources. 

In general, the above timescale is too long for the supermassive black hole scenario to interpret the observed short-term RM variations of some FRB repeaters, which requires a scaled-down $a$ and $l_{\rm cloud}$. The FRB source must be very close to the black hole in order to have the observed rapid RM variability.

\section{Discussion on FRB observations}\label{discussion}

\subsection{FRB 190520B}

FRB 190520B shows the second largest $|{\rm RM}|$ value among all observed FRB sources and a significant RM variation $|\delta{\rm RM}/{\rm RM}|\sim1$ within a few months \citep{Niu22,Anna-Thomas22,Dai22}. 
The observed RM sign reversal of FRB 190520B suggests that its RM variation is mainly due to the change in the magnetic field configuration.

Although an expanding SNR with an unchanged field configuration has been considered to contribute to the long-term RM evolution, e.g., the RM evolution of FRB 121102 \citep{Michilli18,Hilmarsson21}, such a scenario cannot explain the non-monotonic irregularity exhibited by FRB 190520B reported recently \citep{Xu21,Anna-Thomas22,Dai22}. 
Thus, if the large RM and RM variation of FRB 190520B mainly originates from an expanding SNR, the turbulence and the relative motion between the FRB source and the SNR must be taken into account. Based on the discussion in Section \ref{SNR}, 
there are two possibilities to contribute to a large relative RM variation within a few months: 1) The SNR is young with an age of less than a few hundred years. Meanwhile, the SNR shell must be thin compared with its radius and appears significantly anisotropic locally. In addition, the observable secular evolution of RM and DM is expected from such a young SNR, which can be tested by future observations. 2) The SNR could be relatively older, and the typical size of the dense clouds near the SNR is sufficiently small. However, the existence of such small-scale clouds has not been confirmed due to the limited angular resolution of observations \citep{Hennebelle12}.

Another possibility for the large RM and RM variation of FRB 190520B is that it has a companion object, such as a massive/giant star with stellar winds (Section \ref{stellarwind}), or a massive black hole with outflows (Section \ref{BH}). 
Compared with other FRB repeaters with smaller RMs, the companion of FRB 190520B should have stronger stellar winds or black hole outflows, or the companion is very close to the FRB source. 
First, for the stellar wind scenario, in order to explain the large RM of $\sim10^4~{\rm rad~m^{-2}}$ of FRB 190520B, the mass loss rate of the companion stellar wind is required to be $\dot M\sim 10^{-8}M_\odot~{\rm yr^{-1}}$ if the companion is at $\sim 1~{\rm AU}$ distance from the FRB source. Such a large mass loss rate implies that the companion might be a massive main sequence star or a giant star. 
In particular, if the binary orbit is elliptical, a large RM variation would occur near the periastron, and periodic evolution of RM variation is expected \citep{Wang22}.
Alternatively, for the black hole outflow scenario, compared with the FRB repeaters with small RMs, the large RM of FRB 190520B implies that it should be much closer to the massive black hole. However, the mass of the massive black hole cannot be too large, considering that some FRB sources have large offsets from the centers of their host galaxies \citep{Chatterjee17,CHIME20b,Xu21}.

\subsection{FRB 121102}

FRB 121102 has the largest $|{\rm RM}|$ value among all observed FRB sources, and its RM decreased by $|\delta{\rm RM}|\sim |{\rm RM}|\sim10^{5}~{\rm rad~m^{-2}}$ within a few years with the RM sign remaining unchanged \citep{Michilli18,Hilmarsson21}. 
According to Eq.(\ref{eq:deltaRM}), there are two possibilities for its large relative RM variation ($|\delta {\rm RM}/{\rm RM}|\sim1$): 1) It is due to a large relative DM variation, $\delta{\rm DM}/{\rm DM}\sim1$. 
Since the observed DM variation of FRB 121102 is small, $\delta{\rm DM}\sim 1~{\rm pc~cm^{-3}}$ \citep{Hessels19}, the DM of the region contributing to the large RM is required to be also small, i.e. ${\rm DM}\sim\delta{\rm DM}\sim 1~{\rm pc~cm^{-3}}$. 
Thus, the estimated average magnetic field strength is $\left<B_\parallel\right>_L\sim 0.1~{\rm G}$ according to Eq.(\ref{Bfield}), suggesting that the magnetic field near FRB 121102 is extremely strong, much stronger than the observed magnetic fields of most SNRs and pulsar wind nebulae in the Milky Way \citep[e.g.,][]{Reynolds12}. 
2) It is due to the change of the field configuration, $\left|\delta\left<B_\parallel\right>_L/\left<B_\parallel\right>_L\right|\sim1$, which implies that the magnetic field configuration near the source is dynamically evolving during the past few years, similar to the scenario of FRB 190520B \citep{Anna-Thomas22,Dai22}.

The monotonic evolution of RM for FRB 121102 within a few years has been attributed to an expanding SNR with unchanged field configuration \citep{Michilli18,Hilmarsson21}. 
In principle, it could be also due to turbulence, similar to the case of FRB 190520B. The present observation cannot rule out this second scenario. If the RM evolution continues to evolve monotonically over a much longer period of time, the first scenario would be favored. However, if in long term the RM evolution becomes random, one then requires very significant turbulence in the putative SNR surrounding the FRB 121102 source.

At last, due to the largest RM and significant RM evolution of FRB 121102, the RM of this FRB source may also be contributed by the stellar winds from the companion star in a binary system or the outflows from a massive black hole.

\subsection{FRB 180916B}

FRB 180916B exhibits a $\sim$ 16-day periodicity in its burst activity \citep{CHIME20b}, and the periodic activity might be due to orbital motion \citep{Ioka20b,Dai20,Lyutikov20,Zhang20e,Li21d,Wang22}, magnetar precession \citep{Levin20,Zanazzi20,Li21}, or neutron star rotation \citep{Beniamini20,Xu21b}. However, the long-term RM evolution and burst-to-burst RM variations appear to be unrelated to the periodic activity and the activity cycle phase during the three-year observations \citep{Mckinven22}. 
There could be several possible reasons: 1. The periodic activity of radio bursts is due to the precession or rotation of the FRB source, and the long-term RM evolution is due to the SNR expansion. 
Within this scenario, the secular component of the observed $|{\rm RM}|$ would monotonically decrease within a period of time much longer than the current observing time.
2. The periodic activity is due to the precession or rotation of the FRB source, and the long-term RM evolution is due to the orbital motion of the FRB source in a binary system or around a massive black hole. In this case, the orbital period should be much longer than three years, and the periodic evolution of RM could be tested by  future observations over a longer period of time. 3. The periodic activity of radio bursts is caused by orbital motion, but the long-term RM variation of FRB 180916B is due to the intrinsic evolution of the stellar wind or black hole outflow along the line of sight. The variation timescale of the stellar wind or black hole outflow should be much longer than the observed 16-day activity period. 
In this case, the long-term evolution of RM should be random, because the periodic feature of RM would be smeared out by the evolving wind/outflow. Meanwhile, the typical timescale of long-term evolution should be consistent with the activity rate of the stellar wind or black hole outflow.

\subsection{FRB 20201124A}

The RM variation of FRB 20201124A consists of a constant component and an irregular component within over one month \citep{Xu21}. A similar behavior has been observed in  pulsar binary systems \citep{Johnston96,Johnston05}, which indicates an elliptical orbit of the pulsar. 
Since the periodic evolution of RM variations is easier to achieve for an elliptical orbit compared with a circular orbit, if FRB 20201124A is indeed in a binary system with an elliptical orbit, a significant periodic  behavior of its RM evolution is expected from long-term observations.

\section{Conclusions}\label{conclusion}

FRBs are mysterious radio transients with the physical origin still unknown. As cosmological radio transients, their propagating effects (including dispersion, Faraday rotation, temporal scattering, scintillation, depolarization, etc.) are important probes to reveal the physical properties of the astrophysical environments the radio waves propagate through, including the near-source plasma, interstellar medium, and the intergalactic medium. Different from the DM that is mainly contributed by the intergalactic medium, a large absolute RM value of $\gtrsim(10^2-10^3)~{\rm rad~m^{-2}}$ observed at a high Galactic latitude can be only contributed by the magnetized environment near an FRB source. Furthermore, since the interstellar medium and intergalactic medium are not expected to vary in a short time, an observed RM variation can be only attributed to the dynamical evolution or the relative motion of the near-source plasma with respect to the FRB source.

Very recently, some FRB repeaters were found to show significant RM variations \citep{Michilli18,Hilmarsson21,Xu21,Anna-Thomas22,Dai22,Mckinven22}, and the relative variation amplitudes of some repeaters reach $|\delta{\rm RM}/{\rm RM}|\sim1$ within a few months to a few years, e.g., FRB 121102 and FRB 190520B \citep{Michilli18,Anna-Thomas22,Dai22}. The RM variations of FRB repeaters reflect that the near-source environments of FRB repeaters are dynamically evolving (e.g., SNR, stellar flare, etc.) or there is a significant relative motion between the FRB source and the environment (e.g., FRB source in a binary system or in the vicinity of a massive black hole). If the magnetized environment is inhomogeneous, when an FRB propagates in the environment, the electromagnetic waves will be depolarized due to the multi-path propagation effects \citep{Beniamini22,Yang22}, which have been recently confirmed by the observations of some FRB repeaters \citep{Feng22}.

\begin{table*}
\begin{center}

    \caption{Summary of the RM variations for various astrophysical scenarios.}
    \resizebox{\linewidth}{!}{
    \begin{tabular}{cccccc}
    \hline\hline
    \multirow{2}{*}{Astrophysical scenario} 	&\multicolumn{2}{c}{RM} 	&\multicolumn{2}{c}{Random Component} 	&\multirow{2}{*}{DM}\\
    \cline{2-5} 
    &Formula 	&Secular Evolution 	&Timescale Formula$^a$ 	&Variation &\\
    \hline
    Supernova remnant (Section \ref{SNR}) 	&Eq.(\ref{shellRM}) 	&Long-term monotonic 	&Eq.(\ref{SNRvariation}) 	&Random 	&See references$^b$\\
                                        
    \hline
    Companion wind in binary system (Section \ref{stellarwind}) 	&Eq.(\ref{RMw1}-\ref{RMw2}) 	&Periodic 	&Eq.(\ref{windvariation}) 	&Random 	&Eq.(\ref{windDM}) \\
    \hline
    Companion flare in binary system (Section \ref{stellarflare})	&Eq.(\ref{RMflare1}-\ref{RMflare2}) 	&Random 	&Depend on the burst/flare rate$^c$ 	&Random 	&Eq.(\ref{flareDM})\\
    \hline
    Pulsar wind / Pulsar wind nebula / Magnetar flare (Section \ref{pair}) 	&Eq.(\ref{RMp}) 	&Negligible 	&Negligible 	&Negligible 	&Depend on the emission region\\
                                         \hline
    Outflow from massive black hole (Section \ref{BH})	&Eq.(\ref{RMBH}) 	&Periodic 	&Eq.(\ref{variationBH1}-\ref{variationBH2}) 	&Random 	&Eq.(\ref{BHDM})\\
    \hline\hline
    \end{tabular}}\label{table}
    \label{glitch size summary}
\end{center}
$^a$ \scriptsize{Notes: The variation timescale for random component is defined as the typical timescale for $|\delta{\rm RM}/{\rm RM}|\sim 1$.} \\
$^b$ \scriptsize{Notes: The DM evolution for a supernova remnant has been discussed in some previous papers, see \citet{Yang17} and \citet{Piro18}.}\\
$^c$ \scriptsize{Notes: Different from other scenarios, the RM random variation of this scenario is due to the independent random burst/flare rate of FRBs and stellar flares, rather than the turbulence.}\\
\end{table*}

In this work, we have investigated some astrophysical processes that may cause RM variations of an FRB repeater, including SNRs, winds, and flares from a companion in a binary system, pair plasma (pulsar winds, pulsar wind nebulae, and magnetar flares), and outflows from massive black holes, and turbulence induced in these processes. 
The main conclusions for different astrophysical scenarios are shown in Table \ref{table}.
First, we make a general discussion about the statistical properties of random RM variations. We consider that the power spectrum of the RM density ($n_eB_\parallel$) fluctuations satisfies $P(k)\propto k^{\alpha}$, then the RM structure function is $D_{\rm RM}(t)\propto t^{-(\alpha+2)}$ for $v_\perp t<L$ and $D_{\rm RM}(t)\sim\text{constant}$ for $v_\perp t>L$, where $v_\perp$ is the transverse relative velocity between the FRB source and the environment, and $L$ is the outer scale of the inhomogeneous medium. 
During the observing time $t$, the relative RM variation is $|\delta {\rm RM}/{\rm RM}|\sim(v_\perp t/L)^{-(\alpha+2)/2}$ for $v_\perp t<L$ and $|\delta {\rm RM}/{\rm RM}|\sim1$ for $v_\perp t>L$. The measurements of the RM structure function of some FRB repeaters reveal that \citep{Mckinven22} $D_{\rm RM}(t)\propto t^{0.2-0.4}$, leading to $\alpha\sim-(2.2-2.4)$, which implies that the power spectrum of the RM density fluctuations is shallow, and the RM variations are mainly contributed by the electron density fluctuations at small scales. 

On the other hand, secular RM evolution could be attributed to the expansion of a  magnetized shell (e.g., SNR, stellar flares, etc.) or the orbital motion of a binary system (e.g., an FRB source in the stellar winds of a companion in a binary system or in the outflows of a massive black hole). The former scenario predicts that the RM exhibits a long-term monotonic evolution, and the latter scenario suggests that a periodic RM evolution could be detected when the companion has a large-scale strong magnetic field, especially if the orbit is elliptical.

The long-term evolution of the RM contributed by an SNR has been discussed in some previous papers with the assumption that the geometry of magnetic fields along the line of sight keeps unchanged \citep[e.g.,][]{Piro18,Zhao21b}. However, such a model cannot explain the non-monotonic irregular RM evolutions exhibited by some FRB repeaters. In this work, we consider that the medium in an SNR is inhomogeneous due to turbulence and instabilities, and the irregular RM variation is due to the relative motion between the FRB source and the SNR. 
For the FRB repeaters with a large relative RM variation within a few months to a few years, the SNR is required to be young and with significantly anisotropic locally, or the typical size of the nearby dense clouds is extremely small.
For the FRB repeaters with insignificant RM variation, the corresponding SNR is allowed to be much older.

The significant RM variations can also be contributed by the medium from the companion in a binary system. Meanwhile, some evidence suggests that the FRB source might be in a binary system (see a detailed discussion in Section \ref{stellarwind}). When an FRB repeater is in a binary system, the RM variation could be caused by the orbital motion of the binary system or the dynamical evolution of the medium from the companion. 
For a persistent stellar wind, the RM variation is due to the inhomogeneity arising from the turbulence in the anisotropic distribution of the stellar wind. 
In particular, if the binary orbit is elliptical, a large RM variation would occur near the periastron, and periodic evolution of RM variation is expected \citep{Wang22}.

Different from the stellar wind case, stellar flares are the catastrophic release of magnetic energy and are accompanied by CMEs, which are more frequent in low-mass stars. Based on the observed empirical relations of the stellar flares, we calculated the CME RM contribution in a binary system. We found that the RM is almost independent of the CME energy, but is more related to the companion's surface magnetic field and the position where the FRB crosses it. 
Although a large RM value can be generated if the companion is a low-mass star with a strong field and a small separation from the FRB source, a large RM variation is expected during short terms due to the different positions of the CME where the FRB crosses through at different times. The current observation seems not to support such a scenario unless the flares are very frequent. 

In the above discussion, the RM is mainly considered to be contributed by the cold non-relativistic magneto-ionic (ion+electron) plasma. In some astrophysical scenarios, including pulsar winds, pulsar wind nebulae, and magnetar flares, the plasma is composed of relativistic pairs. 
Due to the symmetry of positive and negative charges, the Faraday rotation effect would be canceled, and only the net charges contribute to RM (see Appendix \ref{RMpair} for details). On the other hand, the relativistic motion of electrons would significantly suppress the RM due to the large kinetic mass. Therefore, pulsar winds, pulsar wind nebulae, and magnetar flares cannot contribute significantly to RM and RM variations. This is consistent with observations of most Galactic pulsars and magnetars, especially for the observations of FRB 200428 and radio pulses from SGR J1935+2154.

At last, we discussed the RM contribution by the plasma near a massive black hole. The extremely large RMs with ${\rm RM}\gtrsim 10^4~{\rm rad~m^{-2}}$ have been observed in the vicinity of massive black holes \citep{Bower03,Marrone07,Eatough13}, which have been proposed to be the environment of some FRB repeaters with extremely large RMs. In such a scenario, the random RM variation can be due to the turbulence in the anisotropic distribution of the outflow from the massive black hole or by the interaction between the outflow and nearby clouds. 
Similar to the stellar wind scenario, if the orbit of the FRB source is elliptical, a large RM variation would occur near the periastron, and periodic evolution of RM variation is expected for long-term monitoring. It is worth noting that the mass of the massive black hole cannot be too large in this scenario, because many FRB repeaters were localized at positions far from the centers of their host galaxies \citep{Chatterjee17,CHIME20b,Xu21}.

\section*{Acknowledgements}
We thank the anonymous referee for the helpful comments and suggestions.
We also acknowledge helpful discussions with Shi Dai, Yi Feng, Jonathan Katz, Kejia Lee, Di Li, Dongzi Li, Qiao-Chu Li, Navin Sridhar, Fa-Yin Wang, Wei-Yang Wang, Yuanhong Qu, Zhao-Yang Xia, and Yong-Kun Zhang.
YPY is supported by the National Natural Science Foundation of China grant No. 12003028, the National Key Research and Development Program of China (2022SKA0130101), and the China Manned Spaced Project (CMS-CSST-2021-B11).

\section*{Data Availability}
This theoretical study did not generate any new data.

\bibliographystyle{mnras} 

\appendix

\section{Structure function of RM fluctuations}\label{RMSF}

In this appendix, we calculate the RM structure function following \cite{Lazarian16}.
We define the position on the plane of the sky as $\overrightarrow{x}$ and the distance along the LOS as $s$.
The Faraday RM can be written as 
\be
{\rm RM}(\overrightarrow{x})=\kappa\int u(\overrightarrow{x},s) ds,
\ee
where $\kappa\equiv e^3/(2\pi m_e^2c^4)$, and $u(\overrightarrow{x},s)\equiv n_e(\overrightarrow{x},s)B_\parallel(\overrightarrow{x},s)$ is defined as the RM density. The RM density $u(\overrightarrow{x},s)$ can be described as the sum of its ensemble-average mean and zero mean fluctuations,
\be
u(\overrightarrow{x},s)=u_0+\delta u(\overrightarrow{x},s)~~~\text{with}~\left<\delta u(\overrightarrow{x},s)\right>=0,
\ee
where the subscript ``0'' denotes the mean value, and $\left<...\right>$ is denoted as an ensemble average. The two-point correlation function $\xi_u(l,\Delta s)$ and the structure function $D_u(l,\Delta s)$ of RM density (fluctuations) are described by
\begin{align}
\xi_u(l,\Delta s)&=\kappa^2\left<\delta u(\overrightarrow{x_1},s_1)\delta u(\overrightarrow{x_2},s_2)\right>,\\
D_u(l,\Delta s)&=\kappa^2\left<[u(\overrightarrow{x_1},s_1)-u(\overrightarrow{x_2},s_2)]^2\right>,
\end{align}
where the transverse separation is $l=|\overrightarrow{x_1}-\overrightarrow{x_2}|$, and $\Delta s=s_1-s_2$. Here the statistical homogeneity of the medium is assumed, which is reflected in the fact that $\xi_u(l,\Delta s)$ and $D_u(l,\Delta s)$ only depend on the coordinate difference between the two positions. According to the statistical descriptions presented in \cite{Lazarian16}, we adopt a power-law model of $\xi_u(l,\Delta s)$ and $D_u(l,\Delta s)$,
\begin{align}
\xi_u(l,\Delta s)&=\sigma_{\rm RM}^2\frac{l_{\rm RM}^m}{l_{\rm RM}^m+(l^2+\Delta s^2)^{m/2}},\label{xiu}\\
D_u(l,\Delta s)&=2\sigma_{\rm RM}^2\frac{(l^2+\Delta s^2)^{m/2}}{l_{\rm RM}^m+(l^2+\Delta s^2)^{m/2}},\label{Du}
\end{align}
where $m$ is the scaling slope, $l_{\rm RM}$ is the correlation length of RM density, and $\sigma_{\rm RM}^2=\kappa^2\left<\delta u^2\right>$ is the variance of fluctuations. According to the above equations, the correlation scale $l_{\rm RM}$ can also be defined as $\xi_u(l_{\rm RM},0)=\kappa^2\left<\delta u(\overrightarrow{x}+\overrightarrow{l_{\rm RM}})\delta u(\overrightarrow{x})\right>=\sigma_{\rm RM}^2/2$. 

We consider that the power spectrum of the RM density fluctuations satisfies $P(k)\propto k^\alpha$ for $L^{-1}<k<l_0^{-1}$, where $k=2\pi/l$ is the spatial wavenumber, $L$ and $l_0$ are the outer scale and inner scale, respectively. 
The power spectra with $\alpha<-3$ and $\alpha>-3$ are called the steep spectrum (e.g., $\alpha=-11/3$ for the Kolmogorov scaling) and shallow spectrum, respectively. For the steep spectrum, the fluctuations are dominated by the large scales $\sim L$, which corresponds to the energy injection scale of turbulence, and the correlation scale is $l_{\rm RM}\sim L$. For the shallow spectrum, the fluctuations are dominated by the small scales $\sim l_0$, which is the energy dissipation scale of turbulence, and the correlation scale is $l_{\rm RM}\sim l_0$ \citep{Lazarian16,Xu16}. The relation between the scaling slope $m$ and the spectral index $\alpha$ depends on whether the power spectrum is steep or shallow \citep{Lazarian06,Xu16},
\begin{align}
m&=-(\alpha+3),\alpha<-3,\label{m1}\\
m&=\alpha+3,\alpha>-3,\label{m2}
\end{align}

Next, we define the structure function of RM as
\begin{align}
D_{\rm RM}(\overrightarrow{l})&\equiv\left<[{\rm RM}(\overrightarrow{x}+\overrightarrow{l})-{\rm RM}(\overrightarrow{x})]^2\right>\nonumber\\
&=\left<[\delta{\rm RM}(\overrightarrow{x}+\overrightarrow{l})-\delta{\rm RM}(\overrightarrow{x})]^2\right>.
\end{align}
Notice that the non-standard factor $1/2$ in the definition of $D_{\rm RM}$ in \cite{Lazarian16} has been corrected here due to the standard definition of the structure function adopted. 
According to \cite{Lazarian16}, for the Faraday screen with thickness $\Delta R$, the RM structure function could be calculated by
\begin{align}
D_{\rm RM}(l)=&4\int_0^{\Delta R}(\Delta R-\Delta s)[\xi_u(0,\Delta s)-\xi_u(l,\Delta s)]d\Delta s\nonumber\\
&=4\sigma_{\rm RM}^2\int_0^{\Delta R}(\Delta R-\Delta s)\nonumber\\
&\times\left[\frac{l_{\rm RM}^m}{l_{\rm RM}^m+\Delta s^m}-\frac{l_{\rm RM}^m}{l_{\rm RM}^m+(l^2+\Delta s^2)^{m/2}}\right]d\Delta s.
\end{align}
In order to obtain an analytical resolution of the above integral, we adopt the following approximations: 
1) we are mainly interested in the case of $l\lesssim \min (L,\Delta R)$, because the RMs separated by $l\gtrsim \min (L,\Delta R)$ should be independent, leading to the structure function $D_{\rm RM}(l)\sim\text{constant}$ for $l\gtrsim \min (L,\Delta R)$; 
2) the integrand term $\Delta R-\Delta s$ becomes approximately $\Delta R-\Delta s\sim\Delta R$ for $\Delta s\ll\Delta R$. 
(1) for $l<l_{\rm RM}$, one has 
\begin{align}
D_{\rm RM}(l)&\simeq4\sigma_{\rm RM}^2\Delta R\int_0^{l}\left\{\left[1-\fraction{\Delta s}{l_{\rm RM}}{m}\right]\right.\nonumber\\
&\left.-\left[1-\left(\fraction{l}{l_{\rm RM}}{2}+\fraction{\Delta s}{l_{\rm RM}}{2}\right)^{m/2}\right]\right\}d\Delta s\nonumber\\
&+4\sigma_{\rm RM}^2\Delta R\int_l^{l_{\rm RM}}\frac{ml_{\rm RM}^m\Delta s^{m-2}l^2}{2(l_{\rm RM}^m+\Delta s^m)^2}d\Delta s\nonumber\\
&\simeq4\sigma_{\rm RM}^2\Delta R\left[\int_0^{l}\fraction{l}{l_{\rm RM}}{m}d\Delta s
+\int_l^{l_{\rm RM}}\frac{m\Delta s^{m-2}l^2}{2l_{\rm RM}^m}d\Delta s\right]\nonumber\\
&\sim
\left\{
\begin{aligned}
&\sigma_{\rm RM}^2\Delta Rl\fraction{l}{l_{\rm RM}}{m}, &&m<1,\\
&\sigma_{\rm RM}^2\Delta Rl\fractionz{l}{l_{\rm RM}}, &&m>1.
\end{aligned}
\right.
\end{align}
(2) for $l>l_{\rm RM}$, one has 
\begin{align}
D_{\rm RM}(l)&\simeq4\sigma_{\rm RM}^2\Delta R\int_0^{l}\left(\frac{l_{\rm RM}^m}{l_{\rm RM}^m+\Delta s^m}\right)d\Delta s
\\\nonumber
&\sim4\sigma_{\rm RM}^2\Delta R\left[\int_0^{l_{\rm RM}}d\Delta s+\int_{l_{\rm RM}}^l\fraction{\Delta s}{l_{\rm RM}}{-m}d\Delta s\right]\nonumber\\
&\sim
\left\{
\begin{aligned}
&\sigma_{\rm RM}^2\Delta R l_{\rm RM}(l/l_{\rm RM})^{1-m}, &&m<1,\\
&\sigma_{\rm RM}^2\Delta R l_{\rm RM}, &&m>1.
\end{aligned}
\right.
\end{align}
Some factors with the order of the unit magnitude are discarded in the above approximations. In the following discussion, we are only interested in the case with $m<1$, which is common in most astrophysical turbulence scenarios \citep{Hennebelle12,Lazarian09}.
We must notice that in the above calculation, the power-law model given by Eq.(\ref{xiu}) and Eq.(\ref{Du}) has been used in the total integral range. However, the power-law model reflects robust scaling in the inertial range and is satisfied only in 
$l_0<l<l_{\rm RM}\sim L$ for the steep spectrum ($\alpha<-3$), and $l_0\sim l_{\rm RM}<l< L$ for the shallow spectrum ($\alpha>-3$). 
For a thick Faraday screen with thickness $\Delta R>l_{\rm RM}$ that we are interested in here, the RM structure function is finally given by
\begin{align}
D_{\rm RM}(l)\sim
\left\{
\begin{aligned}
&\sigma_{\rm RM}^2\Delta R l\fraction{l}{l_{\rm RM}}{-(\alpha+3)},&&l_0<l<l_{\rm RM}\sim L,\\
&\sigma_{\rm RM}^2\Delta R l_{\rm RM},&&l>l_{\rm RM}\sim L,
\end{aligned}
\right.
\end{align}
for a steep spectrum ($-4<\alpha<-3$), and 
\begin{align}
D_{\rm RM}(l)\sim
\left\{
\begin{aligned}
&\sigma_{\rm RM}^2\Delta R l\fraction{l}{l_{\rm RM}}{-(\alpha+3)},&&l_0\sim l_{\rm RM}<l<L\sim\Delta R,\\
&\sigma_{\rm RM}^2\Delta R^2\fraction{\Delta R}{l_{\rm RM}}{-(\alpha+3)},&&l\gtrsim L\sim\Delta R,
\end{aligned}
\right.
\end{align}
for a shallow spectrum ($-3<\alpha<-2$) and $L\sim\Delta R$. Notice that in the above equations the condition of $D_{\rm RM}(l)\sim\text{constant}$ for $l\gtrsim \min (L,\Delta R)$ has been used, because the RMs separated by $l\gtrsim\min (L,\Delta R)$ should be independent.

\section{Faraday rotation measure from a relativistic pair plasma}\label{RMpair}

In this appendix, we discuss the RM contribution from a pulsar wind. 
We generally discuss the dispersion relation of a pair plasma not satisfying electric neutrality. For the electromagnetic waves with wavevector $k$ and angular frequency $\omega$, the dispersion relations of right and left circular polarized waves are \citep[e.g.,][]{Stix92}
\begin{align}
\frac{c^2k^2}{\omega^2}&=1-\sum_s\frac{\omega_{ps}^2}{\omega(\omega+\omega_{Bs})}\simeq1-\frac{\omega_p^2}{\omega^2}+\frac{\omega_p^2\omega_B}{\mathcal{M}\omega^3}~~~{\rm for~R~mode},\nonumber\\
\frac{c^2k^2}{\omega^2}&=1-\sum_s\frac{\omega_{ps}^2}{\omega(\omega-\omega_{Bs})}\simeq1-\frac{\omega_p^2}{\omega^2}-\frac{\omega_p^2\omega_B}{\mathcal{M}\omega^3}~~~{\rm for~L~mode},
\end{align}
where $\mathcal{M}$ is the pair multiplicity, $\omega_p^2=\omega_{pe^+}^2+\omega_{pe^-}^2$ is the total plasma frequency, $\omega_{pe^+}$ and $\omega_{pe^-}$ are the plasma frequencies of positrons and electrons, respectively, $\omega_B=\omega_{Be^+}=-\omega_{Be^-}=eB/m_ec$ is the electron cyclotron frequency, and here we assume that $n_{e^-}<n_{e^+}$ and $\omega\gg\omega_B$.

We define the laboratory frame as $K$ and the pulsar wind comoving frame as $K'$. For an electromagnetic wave with frequency $\omega$ and wavevector $k$, the Lorentz transformations of frequency and wavevector between the two frames are
\be
\omega'=\gamma\left(\omega-k_\parallel c\beta\right),~~~~~~~~k_\parallel'=\gamma\left(k_\parallel-\frac{\omega\beta}{c}\right)~~~~~{\rm and}~~~~~k_\perp'=k_\perp,
\ee
where $\beta=v/c$ is dimensionless velocity.
If the wavevector is almost along the line of sight, one has
\be
n'=\frac{n-\beta}{1-n\beta},~~~~~~n=\frac{n'+\beta}{1+n'\beta}.
\ee
Approximately, one has
\be
n^2=\left(\frac{n'+\beta}{1+n'\beta}\right)^2
=1-\frac{1}{\gamma^2(1+n'\beta)^2}(1-n'^2)
\simeq1-\frac{1}{4\gamma^2}(1-n'^2),
\ee
for $n'\sim1,\beta\sim1$ as we are interested in here.
In the $K'$ frame, one has
\be
n_e'\sim \frac{n_e}{\gamma},~~~~~B_\parallel'\sim B_\parallel,~~~~~\omega'\sim\frac{\omega}{2\gamma}.
\ee
We write the wave dispersion relation of the right/left circular wave in the $K'$ frame as
\be
\frac{c^2k'^2}{\omega'^2}\simeq1-\frac{\omega_p'^2}{\omega'^2} \pm \frac{\omega_B'\omega_p'^2}{\mathcal{M}\omega'^3},
\ee
for $\omega'\gg\omega_B'$. 
Notice that for Faraday rotation with polarization angle satisfying $\Delta\phi\propto\nu^{-2}$, the condition of $\omega'\gg\omega_B'$ would be necessary. 
We define $r_c$ as 
\be
\omega_B'(r_c)\sim\omega', 
\ee
and the magnetic field at $r_c$ is
\be
B(r_c)\sim \frac{2\pi m_ec}{e}\left(\frac{\nu}{\gamma}\right)\simeq3.6~{\rm G}\gamma_2^{-1}\nu_9.
\ee
The classical Faraday rotation is available only for $r>r_c$.
The dispersion relation in the laboratory frame is
\be
\frac{c^2k^2}{\omega^2}\simeq1-\frac{1}{4\gamma^2}\left(\frac{4\gamma\omega_p^2}{\omega^2} \mp \frac{8\gamma^2\omega_B\omega_p^2}{\mathcal{M}\omega^3}\right)\simeq1-\frac{\omega_p^2}{\gamma\omega^2} \pm \frac{2\omega_B\omega_p^2}{\mathcal{M}\omega^3}.
\ee
After propagating a distance $d$ from $r_c$, the frequency-dependent polarization position angle is
\begin{align}
\psi&\simeq\frac{1}{2}\int_{r_c}^d\left|k_R-k_L\right|ds\simeq\frac{1}{2c\omega^2}\int_{r_c}^d\frac{2\omega_B\omega_p^2}{\mathcal{M}}ds\nonumber\\
&=\left(\frac{e^3}{\pi m_e^2c^2}\frac{1}{\mathcal{M}}\int_{r_c}^dn_eB_\parallel ds\right)\nu^{-2},
\end{align}
where $\omega\gg\omega_p/\sqrt{\gamma}$ is required for the approximation, which can be easily satisfied in the pulsar wind.
As shown from the above result, only the net charges make a contribution to the Faraday rotation, and the relativistic effect disappears for the Faraday rotation.
According to $\psi={\rm RM}\lambda^2$, the effective RM could be written as
\be
{\rm RM}=\frac{e^3}{\pi m_e^2c^4}\frac{1}{\mathcal{M}}\int_{r_c}^dn_eB_\parallel ds,
\ee
which is suppressed by a factor of $\mathcal{M}/2$ compared with the classical result.
The above result assumes that the plasma is cold in the $K'$ frame. If the plasma is relativistically hot with a typical Lorentz factor $\gamma_{\rm th}$ in the $K'$ frame, the RM contribution is further suppressed by a factor of $\gamma_{\rm th}^2$ due to the relativistic mass $m_e\rightarrow\gamma_{\rm th} m_e$. 
One can finally get
\be
{\rm RM}=\frac{e^3}{\pi m_e^2c^4}\frac{1}{\gamma_{\rm th}^2\mathcal{M}}\int_{r_c}^dn_eB_\parallel ds.
\ee
Therefore, compared with the RM contributed by the non-relativistic magneto-ionic (ions+electrons) cold plasma, the RM from relativistic pair plasma is suppressed by a factor of $\gamma_{\rm th}^2\mathcal{M}/2$.

\bsp	
\label{lastpage}

\end{document}